\documentclass[conference]{IEEEtran}
\usepackage{verbatim}
\usepackage{amsmath}
\usepackage{listings}
\usepackage{xcolor}
\usepackage{setspace}
\usepackage{etoolbox}

\definecolor{mygreen}{rgb}{0,0.6,0}
\PassOptionsToPackage{hyphens}{url}
\usepackage{booktabs}
\usepackage{pdfpages}
\ifCLASSINFOpdf
\else
\fi

\usepackage[font=footnotesize,labelfont=bf]{caption}

\begin{document}
\title{\bf Performance Evaluation and Modeling of HPC I/O on Non-Volatile Memory}

\begin{comment}
\author{\IEEEauthorblockN{Wei Liu}
\IEEEauthorblockA{UC Merced\\
wliu34@ucmerced.edu}
\and
\IEEEauthorblockN{Kai Wu}
\IEEEauthorblockA{UC Merced\\
kwu42@ucmerced.edu}
\and
\IEEEauthorblockN{Jialin Liu}
\IEEEauthorblockA{Lawrence Berkeley National Lab\\
jalnliu@lbl.gov}
\and
\IEEEauthorblockN{Feng Chen}
\IEEEauthorblockA{Louisiana State University\\
fchen@csc.lsu.edu}
\and
\IEEEauthorblockN{Dong Li}
\IEEEauthorblockA{UC Merced\\
dli35@ucmerced.edu}
}
\end{comment}

\newcommand{\email}[1]{\texttt{\small{#1}}}
%\numberofauthors{5}
\author{
  \begin{minipage}{4.0cm}
    \centering
    Wei Liu\\
    \email{wliu34@ucmerced.edu}
  \end{minipage}
  \begin{minipage}{4.0cm}
    \centering
    Kai Wu\\
    \email{kwu42@ucmerced.edu}
  \end{minipage} \\  \\
  \begin{minipage}{4.0cm}
    \centering
    Jialin Liu$^\star$\\
    \email{jalnliu@lbl.gov}
  \end{minipage}
  \begin{minipage}{4.0cm}
    \centering
    Feng Chen$^\dagger$\\
    \email{fchen@csc.lsu.edu}
  \end{minipage}
  \begin{minipage}{4.0cm}
    \centering
    Dong Li\\
    \email{dli35@ucmerced.edu}
  \end{minipage}\\\\
 University of California, Merced \qquad $^\star$Lawrence Berkeley National Lab\qquad$^\dagger$Louisiana State University
}


%Telephone: (800) 555--1212, Fax: (888) 555--1212}


\makeatletter
\patchcmd{\@maketitle}
  {\addvspace{0.5\baselineskip}\egroup}
  {\addvspace{-0.5\baselineskip}\egroup}
  {}
  {}
\makeatother

\maketitle

\begin{abstract}

HPC applications pose high demands on I/O performance and storage capability. The emerging non-volatile memory (NVM) techniques offer low-latency, high bandwidth, and persistence for HPC applications. However, the existing I/O stack are designed and optimized based on an assumption of disk-based storage. To effectively use NVM, we must re-examine the existing high performance computing (HPC) I/O sub-system to properly integrate NVM into it. Using NVM as a fast storage, the previous assumption on the inferior performance of storage (e.g., hard drive) is not valid any more. The performance problem caused by slow storage may be mitigated; the existing mechanisms to narrow the performance gap between storage and CPU may be unnecessary and result in large overhead. Thus fully understanding the impact of introducing NVM into the HPC software stack demands a thorough performance study. 

In this paper, we analyze and model the performance of I/O intensive HPC applications with NVM as a block device. We study the performance from three perspectives: (1) the impact of NVM on the performance of traditional page cache; (2) a performance comparison between MPI individual I/O and POSIX I/O; and (3) the impact of NVM on the performance of collective I/O. We reveal the diminishing effects of page cache, minor performance difference between MPI individual I/O and POSIX I/O, and performance disadvantage of collective I/O on NVM due to unnecessary data shuffling. We also model the performance of MPI collective I/O and study the complex interaction between data shuffling, storage performance, and I/O access patterns. 

\end{abstract}


% \fi
\IEEEpeerreviewmaketitle

\section{Introduction}
\label{sec:intro}

Modern high performance computing (HPC) applications are often characterized with huge data sizes 
and intensive data processing.  For example, the Blue Brain project
aims to simulate the human brain with a daunting 100PB memory that needs to
be revisited by the solver at every time step; the cosmology simulation
studying
Q continuum works on 2PM per simulation. Both of these simulations
require transformation of data representation, which poses high
demands on I/O performance and storage capability.

The emerging Non-volatile Memory (NVM) techniques, such as Phase Change Memory~\cite{ieeemicro10:lee} and STT-RAM~\cite{Dieny08},
offer low-latency access, high bandwidth, and persistency.
Their performance is much better than the traditional hard drives,
and close to or even match that of DRAM. 
The non-volatility and high performance of NVM blur
the line between storage and main memory, hinting at opportunities
to overhaul classical IO system and memory hierarchies. 
Table~\ref{tab:nvm_perf} summarizes the characteristics of different NVM technologies and compares them to traditional DRAM and storage technologies.

\begin{table}[h]
\vspace{-5pt}
        \tiny
        \begin{center}
            \caption{Memory Technology Summary~\cite{NVMDB}}
      \label{tab:nvm_perf}
    \begin{tabular}{|p{0.8cm}|p{1.2cm}|p{1.2cm}|p{1.5cm}|p{1.5cm}|}
     \hline
          & \textbf{Read time (ns)} & \textbf{Write time (ns)} &  \textbf{Read BW (MB/s)} & \textbf{Write BW (MB/s)}   \\ \hline
     DRAM       & 10    & 10 & 1,000 & 900        \\\hline
     PCRAM        & 20-200 & 80-$10^{4}$ & 200-800 & 100-800       \\ \hline
     SLC Flash       & $10^{4}$-$10^{5}$   & $10^{4}$-$10^{7}$ & 0.1 & $10^{-3}$-$10^{-1}$            \\ \hline
     ReRAM			 &  5-$10^{5}$ & 5-$10^{8}$ & 1-1000 & 0.1-1000     \\  \hline
     Hard drive      & $10^{6}$   & $10^{6}$   & 50-120 & 50-120   \\ \hline
     \end{tabular}
     \vspace{-15pt}
      \end{center}
\end{table}

\begin{comment}
Though NVM has very tempting features, to cater these NVM characteristics, new written strategies for software must be prepared. However, there’s no certainty about how it works in application level due to its immature for market and the lack of experiments. This article is aiming to use a DRAM based NVM emulator -- PMBD -- to tests its application level performance under different configuration (different page cache, direct POSIX I/O and highly parallel MPI I/O, MPI Individual I/O and Collective I/O). Plus, we made a model about MPI Collective I/O based on detailed profile data. From these concrete experimental results and model, we could give user a suggestion about how softwares should be written in regrading of optimizing performance in NVM.
\end{comment}

The emergence of NVM has compound impacts on
the existing HPC systems and applications. Given the high performance and non-volatility of NVM, we must re-examine the existing I/O system to properly integrate NVM into it.
Using NVM as a fast storage, the previous assumption on 
the inferior performance of storage, such as disk drives, is not valid any more. 
The performance problem caused by slow storage may be mitigated; 
The performance bottleneck along the I/O path may be shifted from storage to other middle-level system components; 
The existing mechanisms to narrow the performance gap between storage and CPU may be unnecessary and result in undesirable overhead. 

In this paper, we analyze the performance of I/O intensive HPC applications with NVM as the high-speed block device. Given its high compatibility, we anticipate that such a block-based NVM model is likely to become the mainstream in industry (e.g., the recently announced Intel Optane~\cite{Optane}) and be adopted in the near future soon . 
We pose the following questions to gain important insight into the application performance with NVM. 

\begin{itemize}
\item What is the impact of NVM on the performance of traditional page cache? Is it still reasonable to use page cache for NVM-based storage?

\item Comparing MPI individual I/O and POSIX I/O based on NVM, what is their performance difference in the HPC domain? With a high-speed NVM device, 
%would such additional layer bring too much overhead? 
would MPI individual bring too much overhead because it brings one extra layer on top of POSIX I/O?

\item MPI I/O introduces collective I/O techniques to optimize application performance, based on the assumption of poor I/O performance. Is it still valid to use those techniques under the deployment of NVM?
\end{itemize}

To answer the above questions, we use a set of representative HPC applications to evaluate their performance based on Intel's Persistent Memory Block Driver (PMBD)~\cite{linux-pmbd}. We make several findings through our study. 

\begin{itemize}
\item 
The benefits of page cache is diminished with the deployment of NVM, but still plays an important role to improve I/O performance. Comparing with SSD and regular hard drive, 
NVM is less sensitive to page cache size when the working set size of the application is very large. 
This is due to the superior performance of NVM. 
However, when the working set can be accommodated in page cache, NVM does not exhibit significant performance advantages over SSD and hard drive. 
%For example, when 2GB HACC benchmark working on 11GB page cache, performance difference has only $4\%$ difference among three kinds of devices.

\item MPI individual I/O and POSIX I/O have minor performance difference with the existence of NVM. 
The overhead of MPI individual I/O is not pronounced, even if we use NVM as a fast storage.
In a single-node deployment, MPI individual I/O performs only $4.87\%$ worse than POSIX I/O. In a multiple-node deployment, there is almost no performance difference between the two. This indicates that given the current highly optimized implementation of MPI individual I/O, the performance overhead of MPI individual I/O would not become a problem for the future HPC, even if we have a fast storage device, such as NVM.
%even a $25.33\%$ performance loss is introduced due to the network effect. 

\item MPI collective I/O can perform worse than MPI individual I/O with the deployment of NVM. MPI collective I/O aims to aggregate I/O operations to improve performance of MPI individual I/O.  
%However, the benefits of using collective I/O with NVM is reduced, given the good performance of NVM. 
%Further, the I/O aggregation introduces communication overhead, which can overweight the benefits of using collective I/O. 
However, the data shuffling cost in MPI collective I/O is often larger than the performance benefit of collective I/O, given the high speed of NVM. For example, our results show that using collective I/O for a workload with random I/O data accesses from multiple MPI processes performs 38.4\% worse than using MPI individual I/O for the same workload in NVM.
%For an instance, running 16GB IOR benchmark on multiple nodes with 16 processes 4 aggregators, NVM has $38.39\%$ performance lost, while HDD has $57.69\%$ performance boost.

\end{itemize}

Based on our observations, in this paper we further introduce a performance model 
to analyze the tradeoff between I/O aggregation overhead and benefit. Based on the model, we explore how the collective I/O should be employed with the upcoming NVM technology. %Our model achieves good accuracy (\textbf{TODO: give some number here.}).

%[{\bf Add a roadmap paragraph here - Feng}]

\begin{comment}
\subsection{Our Contribution}
Several contributions are made in this paper: (1). We investigate the application level performance impact of page cache by controlling page cache size. Page cache plays an important role in HPC. With page cache, highly used data could be buffered so less cache miss penalty could be achieved. Bigger page cache will lead to more data be buffered and better performance could be reach. As for high speed NVM storage device, we use experiments to find out page cache still play an important role in speed up performance. (2). MPIIO is a highly used parallel computing architecture standard. This parallel computing standard is aim to improve IO speed by synchronously doing multiple IO threads. We use several comparison experiments to check whether this standard still take a great infect in high speed NVM storages. (3). Collective IO is another technology used in HPC to speed up parallel IO performance. Several profiling works are done in this article to check whether collective IO is useful in high speed performance NVM. Furthermore, we made a model to check if this speed up is suitable for any circumstance and in which environment collective IO will take its effect.
\end{comment}

The paper proceeds as follows.
%This article is written in the following structure: 
Section II covers the background.
%including NVM emulator, applications, and preliminary MPI I/O information. 
Section III presents application performance on NVM 
%characteristics of applications 
under various test environments. Section IV introduces our performance model for the MPI collective I/O. We discuss related work and conclude in Sections V and VII, respectively.

\begin{comment}
talk about how we set NVM performance experiments in different page cache and different MPI I/O pattern. Section IV talks about a profile down experiment in MPI I/O, and a model built on this profiling data. Section V gives a detailed list of our relative works, and discuss about the results from these experiments. Finally, in section VI, we give an overall conclusion about experiment results and software suggestions base on these results.
\end{comment}
%paragraph: parallel I/O

\section{Background}
\label{sec:bg}

\subsection{NVM Usage Model}
%MORE NVM BASIC NEEDED

%NVM, represented by Phase Change Memory (PCM) and Spin-transfer-torque RAM (STT-RAM), is a pivotal technology, providing a variety of attractive technical features, such as low power consumption, high endurance, and byte addressability, DRAM-like access speed, disk-like persistence, etc. 
Drawing a blurry line between traditional volatile memory and persistent storage, NVM has at least two usage models. %as follows. 

%The usage model of NVM is under highly debet in prior research [ref. A protected block]. Most studies have indicated at least following basic models: 
(1) {\bf Memory-based Model}. NVM is treated as the regular, byte-addressable %DRAM-based
main memory: NVM is attached to the memory bus in form of DIMMs and directly managed by the memory controller. The NVM space is exposed to the host as part of physical memory address space, which could be directly accessed through {\tt load} and {\tt store} instructions. To bridge the potential performance gap between NVM-only main memory and the traditional DRAM-only main memory, NVM could be paired with a small portion of DRAM to mitigate intensive writes and enhance lifetime.
%is integrated into memory system and display as partially or even entire memory. 
%NVM is regarded similarly to DRAM and provides byte address to IO interfaces for system. 
On one hand, such a memory-based model provides high performance and directly opens many attractive properties, such as byte addressability and persistence, to applications. 
%For example, applications can declare certain in-memory object non-volatile. 
On the other hand, this model introduces high complexity to programmers, especially for handling data integrity and consistency issues upon power and system failure. 
Prior studies, such as Mnemosyne~\cite{Volos11}, CDDCS~\cite{Venkataraman11}, and NV-heap~\cite{Coburn11}, aim to provide an easy and flexible programming interface to alleviate such a programming burden. Also, in order to fully exploit the potential of memory-based model, applications have to be redesigned to fit this model,
which introduces backward compatibility issues. 

(2) {\bf Storage-based model}. Another model is to use NVM as a block device, similar to
traditional HDD or SSD: NVM can be used to directly displace NAND flash in an SSD and managed by an I/O controller. The host can access the device through a regular block I/O interface (e.g., PCI-E or SATA) via {\tt read} and {\tt write} commands. Limited by the I/O bus bandwidth, the storage-based model cannot fully exploit its potential, such as  byte-addressability. However, this scheme provides a maximum compatibility to the existing applications and operating systems, which allows it to be a simple drop-in solution. A user can simply use an NVM device as a regular flash SSD, create partition and file systems atop, and immediately enjoy the high I/O speed. Recently Intel announced their 3D XPoint based product, called Optane, which is a PCI-E device based on the block device model~\cite{Optane}. In this work, 
We assume a storage-based model in this work, which is the most practical NVM solution in the near future. 
%faster device to replace HDDs or SSDs as a persistent storage device. 
%(3). Hybrid Model: NVM is placed between memory and storage, been regarded as a link bridge between the two devices and smoothen the speed difference between each.

%All the three different models have its advantages and disadvantages, however current studies are more likely to choose storage-based model. Performance: According to table I, NVM speed still could not compete with DRAM while it’s much faster than HDD or SSD devices; Protection: According to persistent feature of NVM, the memory-based model has much greater risk of data corruption than the storage model; Ordering Persistent: A power failure could cause data loss, as in volatile CPU cache, in memory-based model. While, the storage-based model provides a write barriers mechanism to safely persist data; Compatibility: Storage-based model provides the best compatibility to existing systems without extra modify for existing architecture and softwares.

\subsection{MPI Collective I/O}
\label{sec:mpi_coll}
%MPI is a highly used parallel high performance computing stander. MPI-IO has a very robust system and widely supported versatility and flexibility of MPI data types. 
In conventional disk based storage, I/O performance is highly sensitive to not only the amount of data being accessed but also the access pattern (e.g., sequential vs. random). In an MPI-based application, multiple I/O streams could be issued individually and independently from multiple MPI processes, which is normally considered as the worst situation for disk drives, because this situation creates a disk head's ``seek storm'' and causes performance loss. Thus, creating a disk-friendly access pattern is an important consideration by MPI I/O. 
%[{\bf I added this part, but not sure if it matches the MPI definition. please verify this statement. -Feng}]

\begin{figure}
\centering
\includegraphics[width=0.48\textwidth, height=0.12\textheight]{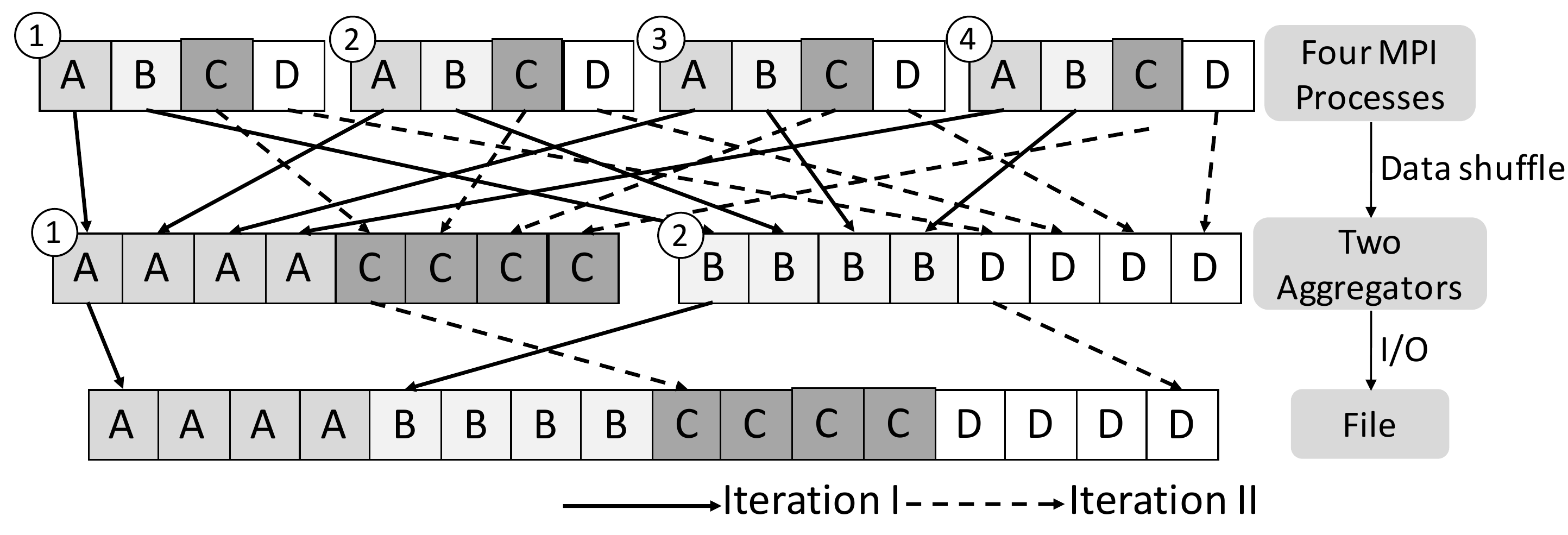}
\caption{The MPI collective I/O scheme. The numbers in circles are MPI process IDs. There are two aggregators (MPI processes 1 and 2) in this example. Letters $A$, $B$, $C$, and $D$ represent data from four contiguous blocks on NVM.}
\label{fig:coll_IO_mech}
\vspace{-15pt}
\end{figure}

Collective I/O is a mechanism to improve MPI-based parallel I/O performance. The basic idea of MPI collective I/O is to scatter and gather data between MPI processes that need to perform I/O operations. Such scatter and gather operations are performed by only a limited number of MPI processes, named as {\it aggregator}. 
%\textbf{In common practice, each node often has one aggregator}.
Each aggregator coalesces I/O requests and iteratively performs I/O operations for all MPI processes or a subset of them.
%that fall into the same node as the aggregator. 
Figure~\ref{fig:coll_IO_mech} depicts the MPI collective I/O scheme for write operation. Read operation happens similarly but in an opposite data path.
In the figure, there are two aggregators (MPI processes 1 and 2). 
Each aggregator gathers data from all MPI processes in two iterations. 
Then each aggregator coalesces the data and writes into persistent storage. 
%As the figure illustrates, 
%in each collective iteration, scattered data pieces are gathered by one or several aggregators and coalesced. Then the coalesced data are read from or written into storage devices through the aggregators. 

%The size of data per collective iteration is limited by a ``collective buffer size''. 
The collective I/O approach reduces the number of I/O transactions, enables contiguous I/O operations, and avoids fetching useless data, effectively improving I/O performance for certain workloads.
%\textbf{TODO: slightly extend the above paragraph to explain how MPI collective I/O iteratively send/receive data.}.
However, MPI collective I/O also brings the so-called ``data shuffling'' overhead, which is associated with the process of data gathering (for write operations) and scattering (for read operations). 
%Such data gathering and scattering is called ``data shuffling''.
%In real-world cases, MPI collective I/O has a detection mechanism to determine whether a collective iteration is worth data shuffling. The ratio (we name it as $\tau$) of data participated in data shuffling to total data depends on workload patterns. 
%[{\bf define tau here, the ratio of what? -- Feng}]

%[{\bf Some descriptions are need to explain the aggregators and processes. The top and bottom part means two aggregator? what does the different shade of the boxes mean? - Feng}]

Given the poor performance of conventional storage devices, the data shuffling overhead is often overweighted by performance benefits of optimized I/O operations from MPI collective I/O. %(see Figure~\ref{fig:coll_IO_mech}). %further explains the MPI collective I/O.
However, with high-speed solid state storage, such as NVM and SSD, which are relatively insensitive to I/O patterns (e.g., random accesses) and deliver much higher I/O performance, MPI collective I/O may not always remain advantageous.
%especially considering the involved data shuffling overhead. 

The current MPI library also allows individual I/O, where MPI processes conduct I/O operations individually without the coordination of MPI collective I/O and do not involve data shuffling. %In this paper, we will particularly study MPI collective and individual I/O. 

\subsection{Benchmarks}
For our experimental study, we have carefully selected four representative I/O intensive HPC benchmarks.

\subsubsection{MADBench2}  This benchmark is a ``stripped-down'' version of MADCAP (a Microwave Anisotropy Dataset Computational Analysis Package)~\cite{site:madbench}. 
%The goal of these experiments is to extract the wealth of cosmology information embedded in the CMB that is related to the universe at an age of about 400,000 years after the Big Bang. 
MADBench2 %is further developed from MADBench2 and could 
%can be run in either ``full'' mode or ``I/O'' mode. I/O mode replaces all computation with dummy work and just measures read and write times. This could be used to test and tune file system performance. 
has an I/O mode that performs MPI I/O in three phases, S, W, and C.
The three phases have complicated write-only, read-only, and read/write operations respectively.

\begin{comment}
The matrix operations are distributed over processors and  build, summed and inverted in phase ``S'', then matrix products are computed in phase ``W'', phase ``C'' reads the results above for an additional likelihood function as well as a quadratic bin-power correction check. Overall, phase ``S'' does write operations only, phase ``C'' does read operations only while phase ``W'' does both.
MADBench2 can be configured with different parameters to change workload size.
\end{comment}
%including ``NO\_PIX'' to specify the number of pixels, ``NO\_BIN'' to specify number of multiple bins, ``NO\_GANG'' to specify  the number of independent work gangs to divide the processors into, and ``BLOCKSIZE'' to specify the size per block used in ScaLAPACK operations.
%[{\bf not sure if we need to give configurations for these benchmarks, somewhere in the paper - Feng}]

\subsubsection{IOR} IOR is a benchmark widely used to study parallel I/O performance at both POSIX and MPI-IO levels~\cite{site:ior}. It is highly configurable and supports various I/O patterns,
including ``sequential'' and ``random offset'' file access,
and individual I/O and collective I/O.
%For example, it supports all ``POSIX'', ``MPIIO'', ``HDF5'', ``NCMPI'' I/O API, both ``individual I/O'' and ``collective I/O'', both ``sequential'' and ``random offset'' file access. It also has configurable features for different workload size.

\begin{figure}
\centering
\includegraphics[width=0.48\textwidth, height=0.09\textheight]{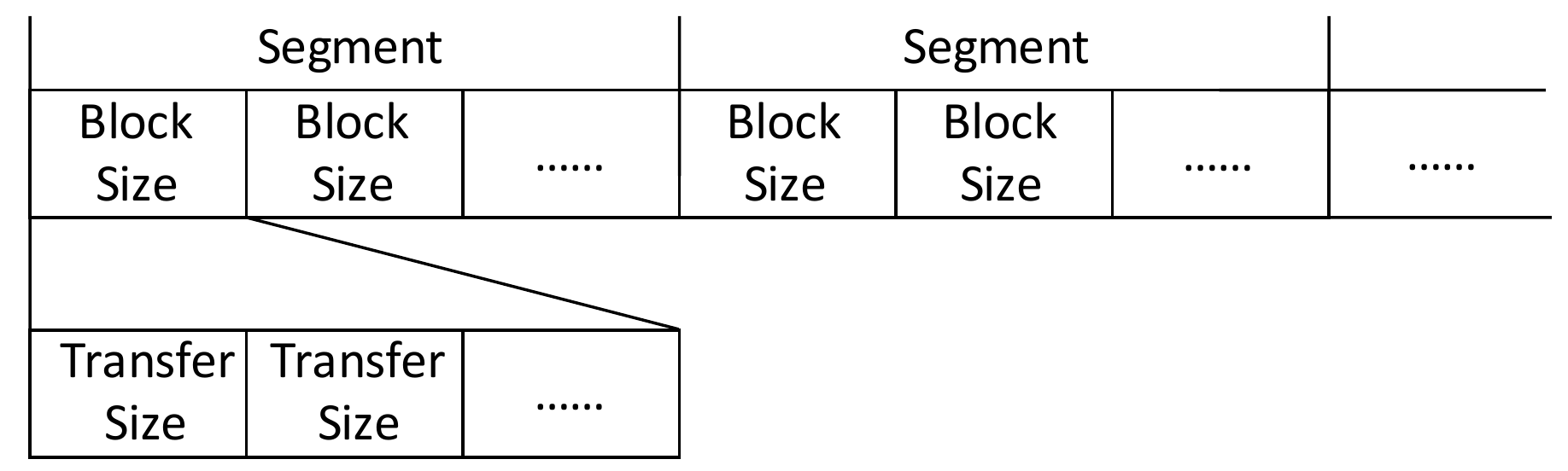}
\caption{Configuration parameters for IOR benchmark.}
\label{fig:ior}
\vspace{-10pt}
\end{figure}

Several IOR configuration parameters are related to our work,
including ``segment count'', ``block size'', and ``transfer size'', shown in Figure~\ref{fig:ior}.
For collective I/O, the given data in an MPI process is partitioned
into segments, and then each segment is further partitioned into blocks. During the data shuffling phase, an MPI process in each iteration of the data shuffling sends or receives at most ``transfer size'' of data. %from another MPI process.
IOR also has a parameter, called ``reorder tasks to random'', 
which enables random I/O accesses. We use this option
for IOR throughout the paper.
\begin{comment}
%The parameter ``block size'' defines a contiguous set of data to write per MPI process; The parameter ``transfer size'' defines the size of transfer in bytes per iteration of MPI collective I/O; 
As Figure~\ref{fig:ior} implies: ``segment count'' and ``block size'' together define how finely the data is divided: ``segment count'' defines how many segments the data will be divided, and ``block size'' further defines how big a data piece should be divided in a data segment; ``segment count'' multiplying ``block size'' will equal to how much data a processor should take control of. Besides, a ``transfer size'' continue to divide the data block into smaller pieces for each parallel I/O iteration.
``reorder tasks to random'' will changes MPI process (task) ordering randomly for I/O operations instead of doing sequentially.
We use this option for IOR throughout the paper.
%\textbf{TODO: what is iteration?}
%\textbf{TODO: what is data segment? Maybe we should explain it in the MPI collective I/O section.}
\end{comment}

\subsubsection{HACC-IO} 
This benchmark is the I/O kernel of HACC (an HPC application based on N-body simulation)~\cite{site:hacc}.
It has random I/O write operations with all-to-all communication patterns.
%HACC benchmark a N-body techniques parallel MPI benchmark simulating the formation of structure in collision less fluids under the influence of gravity in an expanding universe. This benchmark aims at evaluating the performance of I/O system for the hardware by random memory access with all-to-all communication. 
%In this article, to simplify our experiment, we used the I/O kernel of HACC. 
This benchmark allows us to configure the number of particles (``numparticles'') simulated in HACC-IO to change the workload size. The total amount of data to write is the ``numparticles'' multiplied by the number of MPI processes.
%We could set the ``numparticles'' as the number of particles HACC simulating to modify the work load size. The total write our will be ``numparticles'' multiple 38 bytes and number of processes.

\subsubsection{S3aSim.} This benchmark is an MPI-IO based sequence similarity search algorithm framework~\cite{site:s3asim}.
%used to testing and evaluating various I/O strategies. It uses a master-slave parallel programming model with database segmentations. 
S3aSim emulates IO access patterns in mpiBLAST~\cite{site:mpiblast}, which is ``streaming-like'', read-only data accesses. 
S3aSim has five working phases, and we focus on one of the phases (i.e., the I/O phase). 
%This benchmark has totally 5 working phases as ``distribution'', ``computation'', ``gathering results'', ``processing results'' and ``I/O''. We mainly focus on the last phase ``I/O'' to study NVM. 

%[{\bf I revised and moved IOR earlier, so the other two seem not too brief, but I feel we may need to add some extra info to the last two benchmarks - Feng}]

\subsection{PMBD Emulator}
As NVM devices are not available in the market, we use Persistent Memory Block Driver (PMBD)~\cite{MSST14FC}, which is a DRAM based NVM emulator driver, for our experiments. PMBD is a light-weight PM (Persistent Memory) block driver based on an OS kernel module in Linux 2.6.34. It reserves a portion of DRAM-based physical memory space by changing the e820 table in the high memory address space. PMBD provides a standard block I/O interface after being loaded into OS as a regular block device, on top of which partitions and file systems can be created. Internally, the PMBD driver is responsible for mapping the logical block addresses to physical memory pages, receiving the incoming {\tt read} and {\tt write} commands, and translating them to {\tt load} and {\tt store} instructions. From the perspective of application level software and other system components, a PMBD device has no difference from other physical block devices, while it provides configurable features of NVM devices, such as emulating various bandwidths, latencies, protections, etc. 

%Because we assume NVM will be exposed to OS as a contiguous ranged physical device. The mechanism of PMBD driver is mapping NVM physical pages into the kernel virtual address space to make it accessible. For each I/O request, driver will translate read/write request into responding load/store instruction to the physical address mapped.

\subsection{HPC I/O Hierarchy}

\begin{figure}
\centering
\includegraphics[width=0.44\textwidth, height=0.22\textheight]{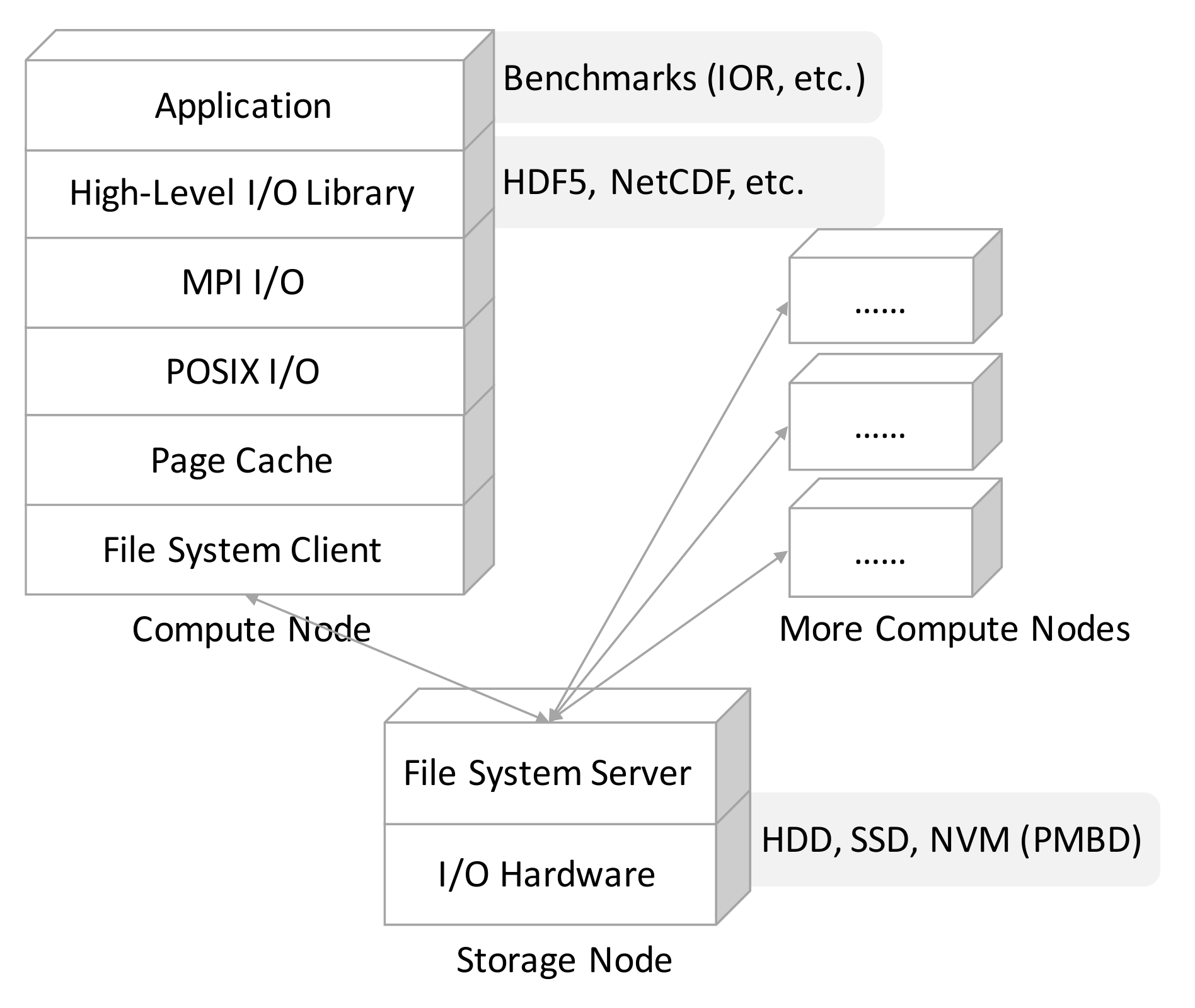}
\vspace{-5pt}
\caption{HPC I/O system Hierarchy}
\label{fig:hie}
\vspace{-20pt}
\end{figure}
%(\textbf{TODO: page cache should be on compute side, IO hardware on storage side -Liu. In Figure 2, page cache can be moved up above the file system as an indivdiual layer, since we need to analyze the page cache effect, which will also fix the problem - Feng})

%This section explain a overview about our system hierarchy, as Figure~\ref{fig:hie}. In the bottom is the I/O storage hardwares. As we want to test, NVM (emulated by PMBD) is one kind of them. And we added HDD, SSD as the comparison. Above I/O hardware (through hardware page cache), parallel file system (NFS in our experiments) connected multiple clusters. And then is the high-level library (we used open-mpi, MPICH, etc.) connected between application (HACC-IO, MADBench2, IOR, etc.) and I/O Middle-ware (MPI-IO, POSIX, etc.).

The I/O stack in a typical HPC system has multiple layers, shown in Figure~\ref{fig:hie}.
%creating a so-called I/O hierarchy (see Figure~\ref{fig:hie}). 
The block devices at the bottom level provide data persistence.
%determine the essential speed of referencing data from persistent storage. 
Given the variety of different storage devices (e.g., HDD, SSD, PMBD), raw data access latencies range from microseconds to milliseconds, and are sensitive to distinct access patterns (e.g., sequential vs. random). To alleviate the impact of slow I/O operations, the page cache layer in the operating system attempts to hold the workload's working set in memory, satisfying most data accesses in DRAM. Due to its ``filtering'' effect, the page cache can have a strong impact on I/O performance. 
%In our experiments, we will particularly study the impact of page cache. 
The file system layer is responsible for managing storage devices and provides a file system abstraction to allow applications to access storage devices, either connected locally or remotely. In our study, we use network file system (NFS) for remote storage access.
%have tested with both Network File System (NFS) and local file systems. 
MPI I/O built on top of POSIX I/O enables coordinated and remote I/O accesses for MPI processes.

%[{\bf Intro to MPI-IO library layer still needs to be added here - Feng}]
%{\bf [Do we need a figure to show the cluster's network topology for experiments?? - Feng]}

\section{Performance Study} 
\label{sec:perf_study}
%To simulating the environment NVM actually will meet in market, we carried out several really applications which test the performance in different configuration.
We present our performance analysis results in this section.
%\subsection{Targeted supercomputer}
%\subsection{Test Environment}
%All our tests are done under Alpha supercomputer hosted at PASA, UC Merced laboratory. Detailed information are shown in Table
We deploy our tests in a local cluster. Each node of the cluster has
two Intel Xeon E5-2630 processors (2.4GHz) with 32GB DDR memory. All nodes in the
cluster are connected through 1Gb Ethernet interconnect. 
We use three types of block devices: one is a regular hard drive (Seagate Constellation.2 500GB hard drive attached by SATA, notated as ``HDD'' in this section), one is an SSD (Intel SSD730 240GB attached by SATA, notated as ``SSD'' in this section), and the third is an NVM device emulated with PMBD. NVM is configured with the same bandwidth and latency as DRAM.
We use MPICH-3.2 for MPI throughout the paper.

\subsection{Impact of Page Cache}
The page cache is a transparent cache for pages originating from a secondary storage device. The operating system (OS) keeps a page cache, which enables quicker accesses to those frequently accessed pages and improves performance.
%the contents of cached pages and overall performance improvements. 
%Testing benchmarks under different page cache settings and comparing the their bandwidth under different settings could give us a hint about how page could affect application performance in different devices.
We measure the performance of the three I/O devices with different page cache configurations and study the impact of the page cache on the observed application performance. 

%By limiting the page cache setting in Linux kernel we could simulating different page cache circumstance. In our experiments, we set different page cache ranged from 1GB to 11GB, which most involves current supercomputing configurations.

%The benchmarks chosen in our experiments are similar or real applications. In this subsection, we chose three real application simulating benchmark: HACC-IO-Kernel, MADBench2, s3salm as our target tests. All benchmarks are compiled and run Open-MPI 1.10. Only one node involved with 4 processors.

\begin{figure}[t]
\centering
\includegraphics[width=0.48\textwidth, height=0.20\textheight]{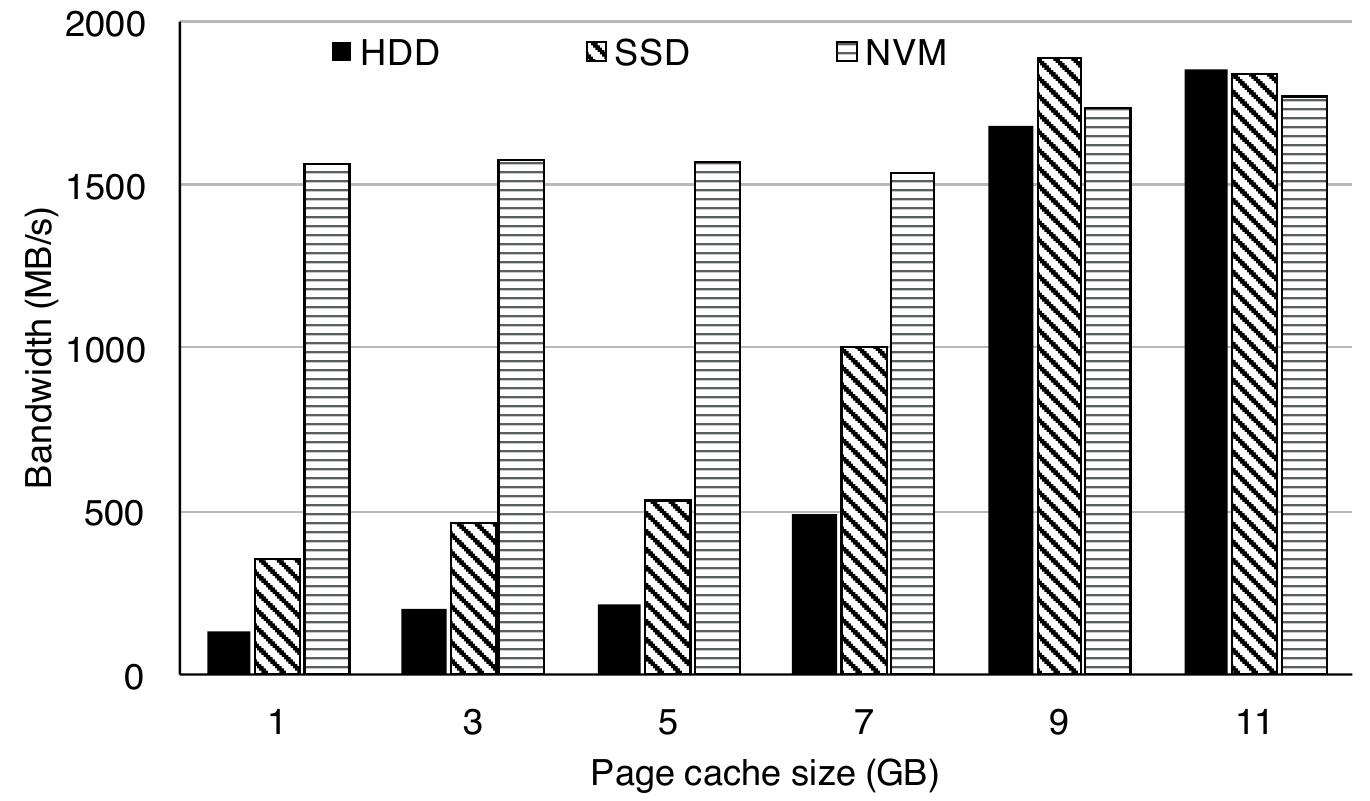}
\vspace{-5pt}
\caption{The performance study for the impacts of page cache on HACC-IO.}
\label{fig:HACC_pagecache}
\vspace{-10pt}
\end{figure}

\begin{figure}[t]
\centering
\includegraphics[width=0.48\textwidth, height=0.20\textheight]{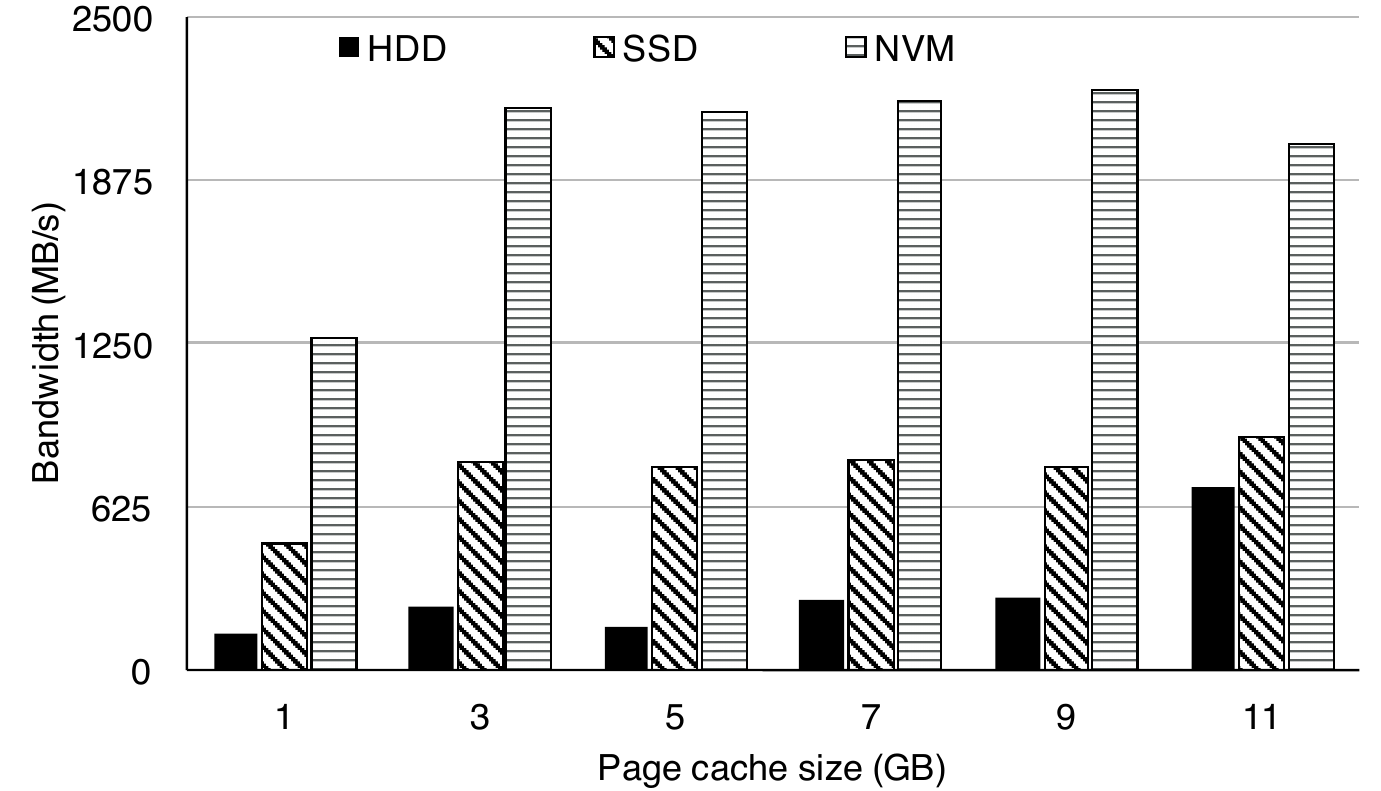}
\vspace{-5pt}
\caption{The performance study for the impacts of page cache on MADBench2.}
\label{fig:MADB_pagecache}
\vspace{-20pt}
\end{figure}

\begin{figure}[t]
\centering
\includegraphics[width=0.48\textwidth, height=0.20\textheight]{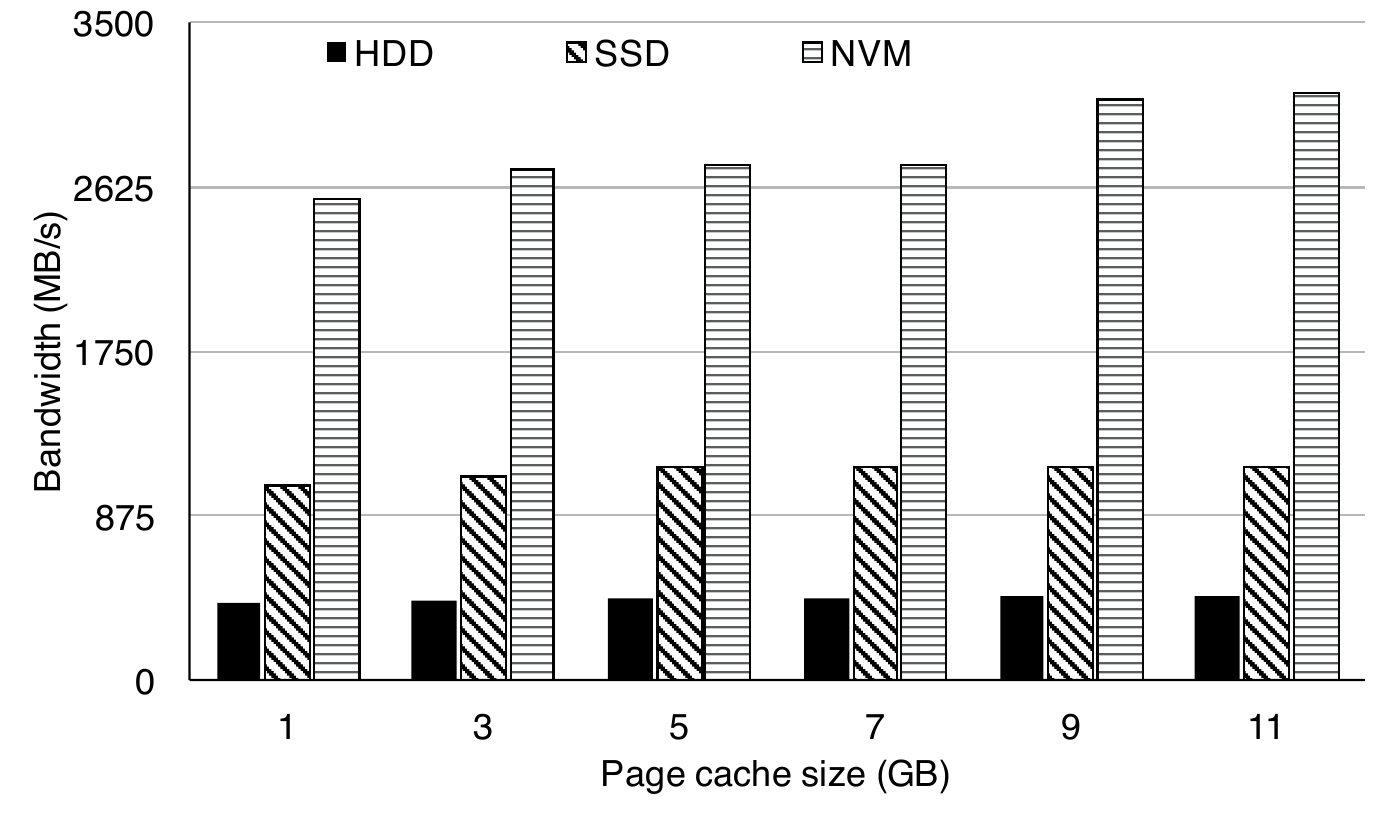}
\vspace{-5pt}
\caption{The performance study for the impacts of page cache on S3aSim.}
\label{fig:SAL_pagecache}
\vspace{-20pt}
\end{figure}

We use three benchmarks in our tests, HACC-IO, MADBench2, and S3aSim. The benchmarks are compiled with gcc 4.4.7 and Open MPI-1.10.0. 
We use one node with four MPI processes for our tests.
Figures~\ref{fig:HACC_pagecache},~\ref{fig:MADB_pagecache}, and~\ref{fig:SAL_pagecache} show the results for HACC-IO, MADBench2, and S3aSim, respectively. 

HACC-IO in Figure~\ref{fig:HACC_pagecache} simulates 13,107,200 particles in total (i.e., numparticles=13,107,200). It computes and then generates about 2GB data for four MPI processes, and writes them into the three block devices. 
%For HACC, totally 2GB data is computed and write into the device we want to test. Time and bandwidth are reported by benchmark. 
%To ensure there's no infect about memory buff, system reboot between each tests. 
%Comparing the result from chart above,  we could found out page cache indeed plays an important role in HDD and SSD, while has little impact in PMBD. When page cache is large enough (like in 11GB and 9GB), benchmark performance is roughly the same. That may because page cache is big enough to buffer all data. Performance drop significantly in HDD and SDD when become smaller, while performance dropped but slightly in PMBD. Over all, from 11GB to 1GB, performance dropped 92.66 percentage, 84.84 percentage, 11.49 percentage in HDD, SSD, PMBD representatively.
The figure reveals that the page cache plays an important role to improve performance for HDD and SSD, while it has a limited impact on the performance of NVM. 
When the page cache size is large (e.g., 9GB and 11GB), there is almost no performance difference between the three devices, because most of the I/O data is cached in the page cache. However, as we reduce the page cache size, there is significant performance difference between the three devices.
%since the 7GB of the page cache.
In general, decreasing the cache size from 11GB to 1GB, the performance of this workload on HDD and SSD is reduced by 92.7\% and 84.8\% respectively, while the performance with NVM is  only  reduced by 11.5\%. This example illustrates well that with high-speed NVM, the effect of the page cache is weakened. 

MADBench2 in Figure~\ref{fig:MADB_pagecache} uses a working set size of about 4GB (particularly the parameters NO\_PIX, NO\_BIN, NO\_GANG, and BLOCKSIZE of MADBench2 are set as 5000, 8, 1, and 1024 respectively), 
%``NO\_PIX'': number of pixels is 5000, ``NO\_BIN'': number of bins is 8, ``NO\_GANG'' number of independent work gang is 1, ``BLOCKSIZE'' bytes block size is 1024), 
larger than that of HACC-IO. 
%MADBench2 has three execution phases, $S$, $W$, and $C$. 
The figure presents the performance of the phase $W$, which includes both read and write operations. 
MADBench2 tells us a story slightly different from HACC-IO. 
As we decrease the cache size from 11GB to 3GB, the performance on the three devices
remains stable. This is because of the fact that MADBench2 has a larger working set size and the page cache is unable to effectively cache all data, including those for MADBench2 and system. However, NVM performs the best among the three cases due to its high bandwidth. 
%Because MADBench2 has a larger working set size, MADBench2 on all of the three block devices has performance loss, 
%since the page cache is unable to effectively cache all data, including those for MADBench2 and system. However, NVM performs the best among the three cases due to its high bandwidth. 

%For MADBench we tried to increase workload size into 4GB. There are three phases in MADBench (S, W and C). We focus on the phase W which both has read and write manipulation. Test latency is reported by benchmark. Bandwidths are got using workload size divide by latency. Figure~\ref{fig:MADB_pagecache} shows the result for our tests. 

%Same as HACC, we found page cache still plays a significant role for MADBench in HDD and SSD, while has a relatively small impact in PMBD. A little different from above is that speed difference has been shown from 11GB case, that because workload size is larger and memory buffer could not take its advantage in this case. Over all, from 11GB to 1GB, performance dropped 79.48 percentage, 45.15 percentage, 37.08 percentage in HDD, SSD, PMBD representatively.

S3aSim in Figure~\ref{fig:SAL_pagecache} uses a working set size of 2GB (with 100 total query number, max size of each query as 5,000, and max count of each query as 10,000). Comparing the performance
of MADBench2 and S3aSim, we find that they have the same performance trend: the NVM has the best performance in all cases.
%but shows different performance when the page cache size is small (1GB). 
But when the page cache is reduced from 3GB to 1GB, 
MADBench2 on NVM has significant performance reduction, $40.16\%$, 
%when the page cache size reduces from 3GB to 1GB; for S3aSim, the performance degradation is only $5.91\%$.
while S3aSim on NVM has only 5.91\% performance reduction.
We attribute such difference in the performance reduction to the distinct data access patterns of the two applications: S3aSim has streaming-like access pattern, hence the page cache cannot work well, no matter how large the page cache size is; for MADBench2, the page cache takes effect, although the caching effect of page cache becomes weaker, when the page cache size is small (1GB).

%Continue our experiment in s3salm. 2GB workload size used here. Test latency reported by benchmark. Bandwidths are compute by size and time. Figure~\ref{fig:SAL_pagecache} shows the result for our tests. 

%As a different story compare to the two experiments above, s3salm is not a little sensitive to page cache in any three kind of devices. Though bandwidths in each device varies a lot, they are very consistent throughout page cache changing. 

\textbf{Conclusions.}
With the emergence of NVM, the impact of page cache on application performance is diminishing. Compared with the traditional HDD and SSD, NVM is relatively insensitive to the page cache size. 

Our study has an important implication on how much page cache space should be allocated for future NVM-based HPC systems. In general, NVM makes it possible to use a smaller page cache, which would save cost and incur ignorable performance impact. We could even explore the possibility of completely bypassing page cache for certain workloads on NVM-based block device, which will save the limited page cache space for other system data and in turn improve the performance of the whole system.
%Summarize the three experiments above, PMBD is not as sensitive to page cache as slow device (HDD, SSD) in certain cases. Though, depending on application I/O attribute all three kind of devices are may not sensitive to page cache, NVM still have a huge market potential in the future.

\subsection{POSIX I/O and MPI Individual I/O}
\label{sec:posix_vs_individual}
%This subsection talks about performance difference in POSIX and MPIIO. We used IOR benchmark, which is a highly used parallel computing benchmark and supporting both POSIX and MPIIO interfaces. Also, to exclude network effect, all experiments are done in single node. The storage node memory is set to 12GB. 
MPI I/O is built on top of POSIX I/O (see Figure~\ref{fig:hie}), and is designed to 
%MPI I/O is meant to 
improve the performance of POSIX I/O in the setting of
parallel I/O and provide user-friendly I/O abstract. %by \textbf{TODO: xxx}.
In the system stack, MPI I/O layer ensures data validness
for MPI I/O operations and re-organizes data distribution for better performance. 
However, as an additional layer in the system stack, MPI I/O could introduce certain overhead. 
%Such a new layer is necessary to ensure data validness
%for MPI I/O operations and re-organize data distribution for better performance. 
%Such overhead, which may not be obvious when using traditional HDD and SDD, may be more pronounced when using NVM, because NVM alleviates
%performance bottleneck at I/O devices and makes the overhead in the other
%system components more obvious.
With conventional disk storage devices, such overhead is negligible compared to its advantages, however, it could be more pronounced with NVM, because NVM alleviates performance bottleneck at I/O devices and makes the overhead in the other system components more obvious.
In this section, we study the performance of MPI individual I/O, and further study the performance of MPI collective I/O in the next section. %[{\bf revised some, but we need to consider if this claim can be supported by the data below -Feng}]

\begin{figure}
\centering
\includegraphics[width=0.46\textwidth, height=0.19\textheight]{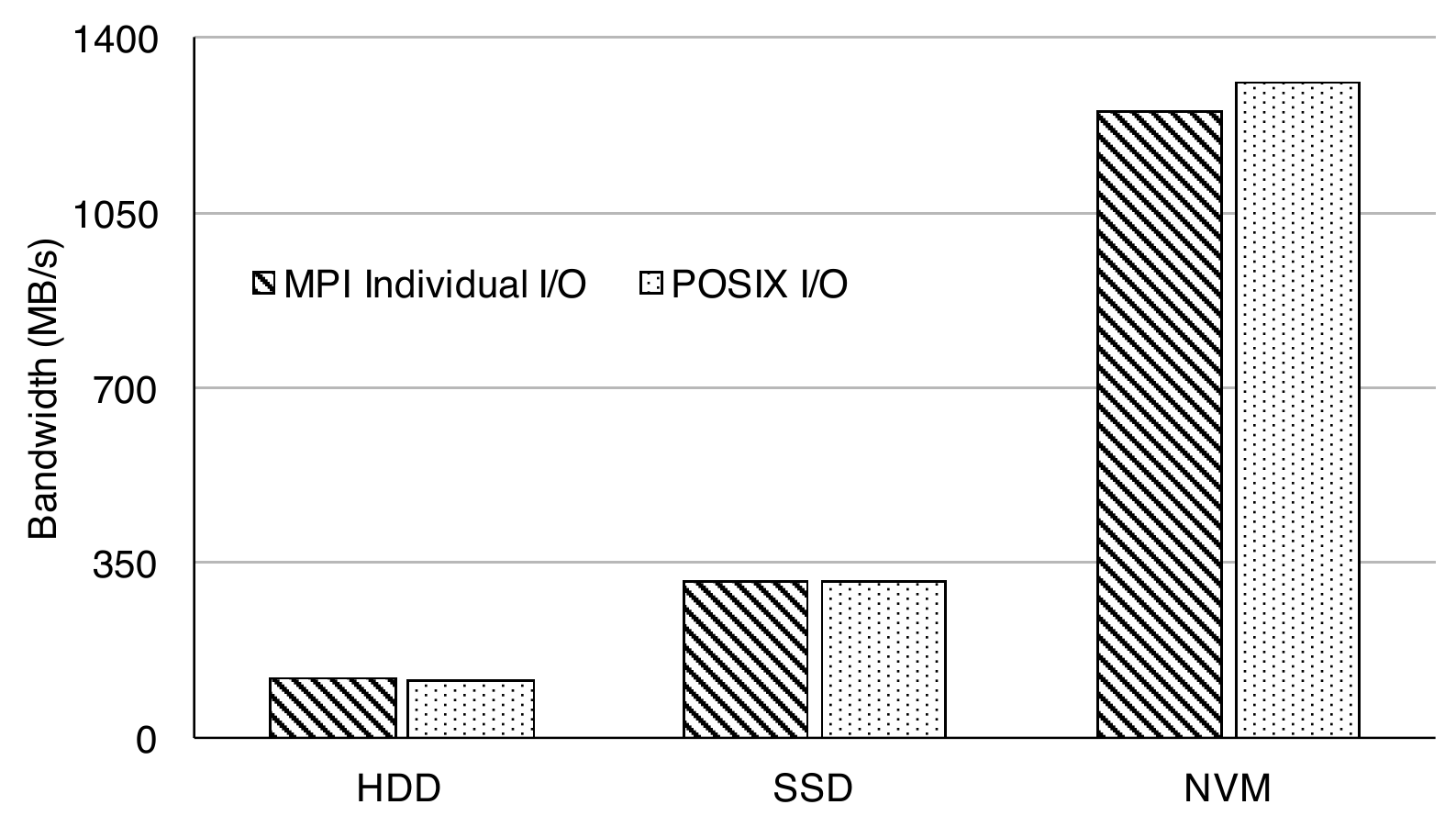}
\vspace{-5pt}
\caption{Comparing the performance of MPI individual I/O and POSIX I/O on a single node with IOR.}
\vspace{-15pt}
\label{fig:IOR_MP_S}
\end{figure}

We first study the performance of POSIX I/O and MPI individual I/O without the involvement of network communication. In particular, we run the IOR benchmark on a single node. We use 4 MPI processes, each of which performs I/O operations. 
For the IOR benchmark, we set ``block size'' as 256MB, ``segment count'' as 2,
and ``transfer size'' as 16MB, and enable ``reorder tasks to random''. The final aggregated result file from IOR is a 16GB file (each MPI process writes 4GB data). 
%We set transfer size as 16MB, so each collective I/O call transfers 16MB in total. 
%Task ordering set to random ordering instead of sequential, so collective I/O will take effect. 
%\textbf{TODO: rephrase the above sentence to explain what does block size, transfer size and task ordering mean.}
Figure~\ref{fig:IOR_MP_S} shows the results. 
%During the experiments, we found there's a huge speed gap between SSD and NVM. To compensate this difference, we introduce an other device -- BigSSD, aside HDD, SSD, NVM, as a speed transition between SSD and NVM. This BigSSD was done by merge two SSDs into one big block device by RAID 0. RAID 0 (also known as a stripe set or striped volume) splits ("stripes") data evenly across two or more disks, without parity information, redundancy, or fault tolerance. The merged big SSD block will have a better performance than each single one.

%As for configuration of IOR, workload size is 16GB as so a little larger than memory size (12 GB). We used several processes (single, 2, 4, 8 till to 64 processes). Block size set as large as 256MB, transfer size set to 16MB, set task ordering to random ordering for read back. Figure~\ref{fig:IOR_MP_S} shows the result for 4 processors situation.

The figure reveals that there is almost no performance difference between MPI individual I/O and POSXI I/O on a single node for HDD and SSD. 
However, when we use NVM, we notice that POSIX I/O performs slightly better than MPI individual I/O by $4.87\%$.
%(\textbf{\% performance difference}).
We attribute the appearance of such performance difference to
the better performance of NVM which makes the overhead of MPI I/O more pronounced.
%[{\bf I think this small performance difference cannot support the argument here. -Feng}]

% BigSSD part
%To verify the above conclusion, we introduce a new system setting that includes two SSD combined as a big block device based on RIAD 0. Such system setting is labeled as ``BigSSD'' in Figure~\ref{fig:IOR_MP_S}. BigSSD increases I/O bandwidth by distributing data on two SSD.Hence, BigSSD has better I/O bandwidth than SSD and HDD and provides better performance. But BigSSD still has smaller I/O bandwidth than NVM. With the deployment of BigSSD, we also notice POSIX I/O performs slightly better than MPI individual I/O by $5.41\%$. Again, such noticeable performance difference is because of better performance of BigSSD.

\begin{figure}
\centering
\includegraphics[width=0.46\textwidth, height=0.19\textheight]{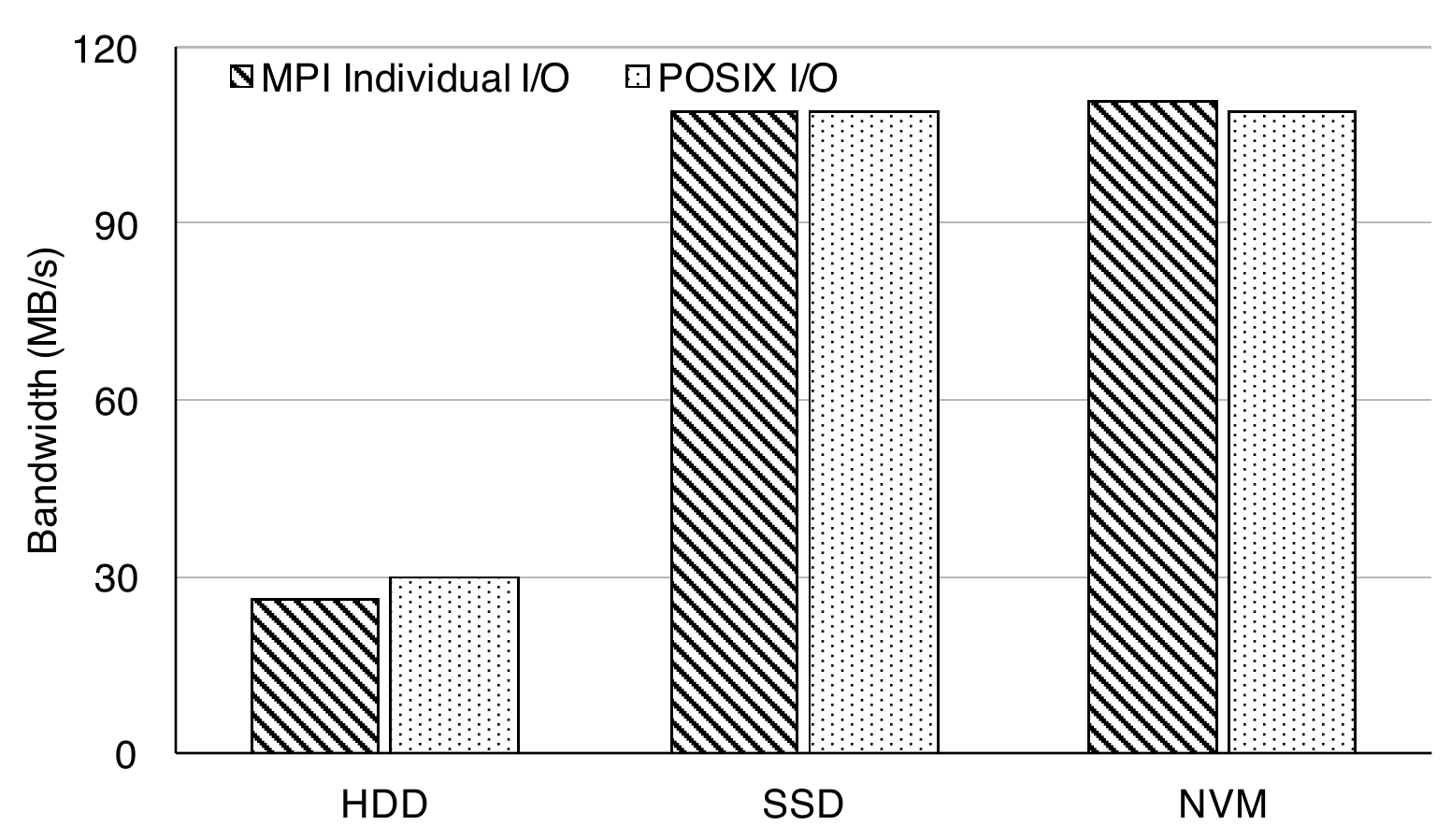}
\vspace{-5pt}
\caption{Comparing the performance of MPI individual I/O and POSIX I/O  on multiple nodes with IOR.}
\vspace{-20pt}
\label{fig:IOR_MP_M}
\end{figure}

To further study the performance of MPI individual I/O and POSIX I/O, we use five nodes and re-do the tests. 
Among the five nodes, four nodes run the IOR benchmark with 4 processes per node (16 processes in total), and the fifth node works as a storage node where the other four nodes remotely perform I/O operations. 
Hence, different from Figure~\ref{fig:IOR_MP_S}, %the tests on a single node, 
such a deployment has the involvement of communication between the four nodes and the storage node. POSIX I/Os are performed with NFS in our test environment.
Figure~\ref{fig:IOR_MP_M} shows the results.

The figure reveals that MPI individual I/O has almost no performance difference than POSIX I/O in all cases,
%leads to better performance than POSIX I/O in all cases, 
no matter whether we use HDD, SSD, and NVM. 
%It seems that MPI individual I/O does better job to optimize I/O performance than POSIX I/O when we have communication.
The communication cost in our tests is the major performance bottleneck, 
%[{\bf I changed from ``one of the major'' to ``the major''. If it is inaccurate, please reverse the change. -Feng}]
much larger than those caused by MPI individual I/O overhead.
Hence, the overhead for MPI individual I/O is not clearly spotted
in the figure, even if we use a fast storage device, such as SSD and NVM.  
%(\textbf{TODO: Can we explain better on why MPI individual I/O has better performance in this case?})

%[{\bf The results here need a better explanation - the result is not strong enough to claim (1) MPI layer causes overhead compared to POSIX IO. 0.88\% difference is very small. (2) Using a faster device makes the overhead more pronounced. Why BigSSD shows a bigger difference (5.41\%) than NVM? -Feng }]

%Actually, from the chart above, we found there's no much different between MPI-IO and POSIX IO in any kind of device. 

%Also, we continue our tests in multiple nodes. In this case, we have one main storage node and 4 remote computing nodes. And we set totally 16 process, each computing node have 4 processes. Figure~\ref{fig:IOR_MP_M} shows the result for this experiment.

%Roughly, the results lead to the same conclusion as single node. In each kind of device, POSIX I/O and MPIIO have roughly the same performance. We have to point out, because network has a big influence between multiple nodes, MPI-IO will has a slightly bigger influence in fast device. As shown in chart, NVM has a certain deduction in performance.

\textbf{Conclusions.}
The emergence of NVM brings better performance, and also may make some overhead more pronounced than before. 
In this section, we study the overhead of MPI individual I/O. 
We find such overhead only sightly impacts performance in a deployment of a single node, and in a multi-node environment, MPI individual I/O has ignorable performance overhead, even if we use NVM.
This finding implies that the current implementation of MPI individual I/O is quite efficient, which would introduce little overhead for the future HPC system equipped with NVM.

%Such overhead is not obvious before the birth of NVM. The performance difference between POSIX I/O and MPI individual I/O because of MPI I/O overhead is such an example.
%We expect that with the deployment of NVM in future HPC, the future I/O performance optimization target may have to be expanded from I/O-only to other system components in the whole system stack.

\subsection{MPI Collective I/O and MPI Individual I/O}
\label{sec:coll_io_perf_study}
%MPI-IO introduced collective I/O to improve the performance in big data piece read and write. Studies shows that collective I/O could helps to reduce the time consumption in network. But most of the studies were done in slow storage device, like HDD. Is collective I/O still has a big performance improvement in fast devices in NVM?

%For this section, we still use IOR benchmark because itself already supports both individual I/O and collective I/O. As the same as previous experiment, both workload size and transfer size should be big enough, so that collective I/O could take effect. Memory size set to 12GB, workload size set to 16GB, block size is 256MB, transfer size is 16MB. We tested our experiment by 4, 16 and 64 processors. There are 1 main storage node connected with 4 remote computing nodes. So, each single computing node has 1, 4 and 16 process representatively. One aggregator for every computing node responses for data shuffle. Figure~\ref{fig:IOR_IC4}, ~\ref{fig:IOR_IC16} and ~\ref{fig:IOR_IC64} shows the results for this part.

MPI collective I/O can bring performance benefit over MPI individual I/O, when I/O operations from MPI processes are interleaved and scattered.
By coalescing I/O operations and reorganizing data between MPI processes, MPI collective I/O can reduce the number of I/O transactions and avoid fetching useless data.
However, this happens at the cost of data shuffling operations between MPI processes, as discussed in Section~\ref{sec:mpi_coll}. 
The design of MPI collective I/O is based on a fundamental assumption that the I/O block device is slow and pattern sensitive, such that the data shuffling cost can be overweighted by the performance benefit of using MPI collective I/O. In this section, we study the performance of collective I/O with NVM, and compare the performance of MPI collective I/O and MPI individual I/O. 

\begin{figure}
\centering
\includegraphics[width=0.46\textwidth, height=0.20\textheight]{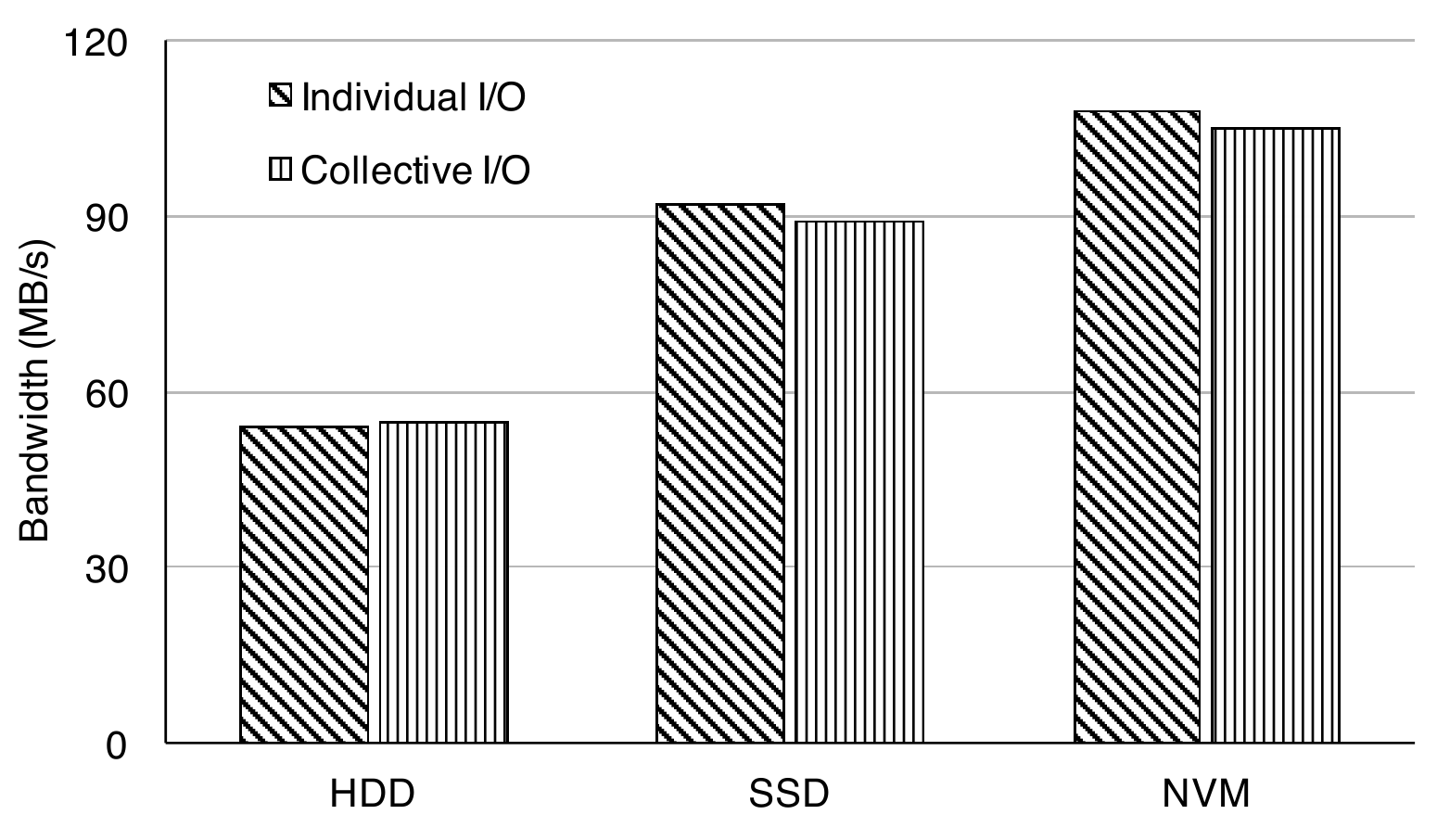}
\vspace{-5pt}
\caption{Comparing the performance of MPI collective I/O and MPI individual I/O (1 process per node) with IOR.}
\vspace{-10pt}
\label{fig:IOR_IC4}
\end{figure}

\begin{figure}
\centering
\includegraphics[width=0.46\textwidth, height=0.20\textheight]{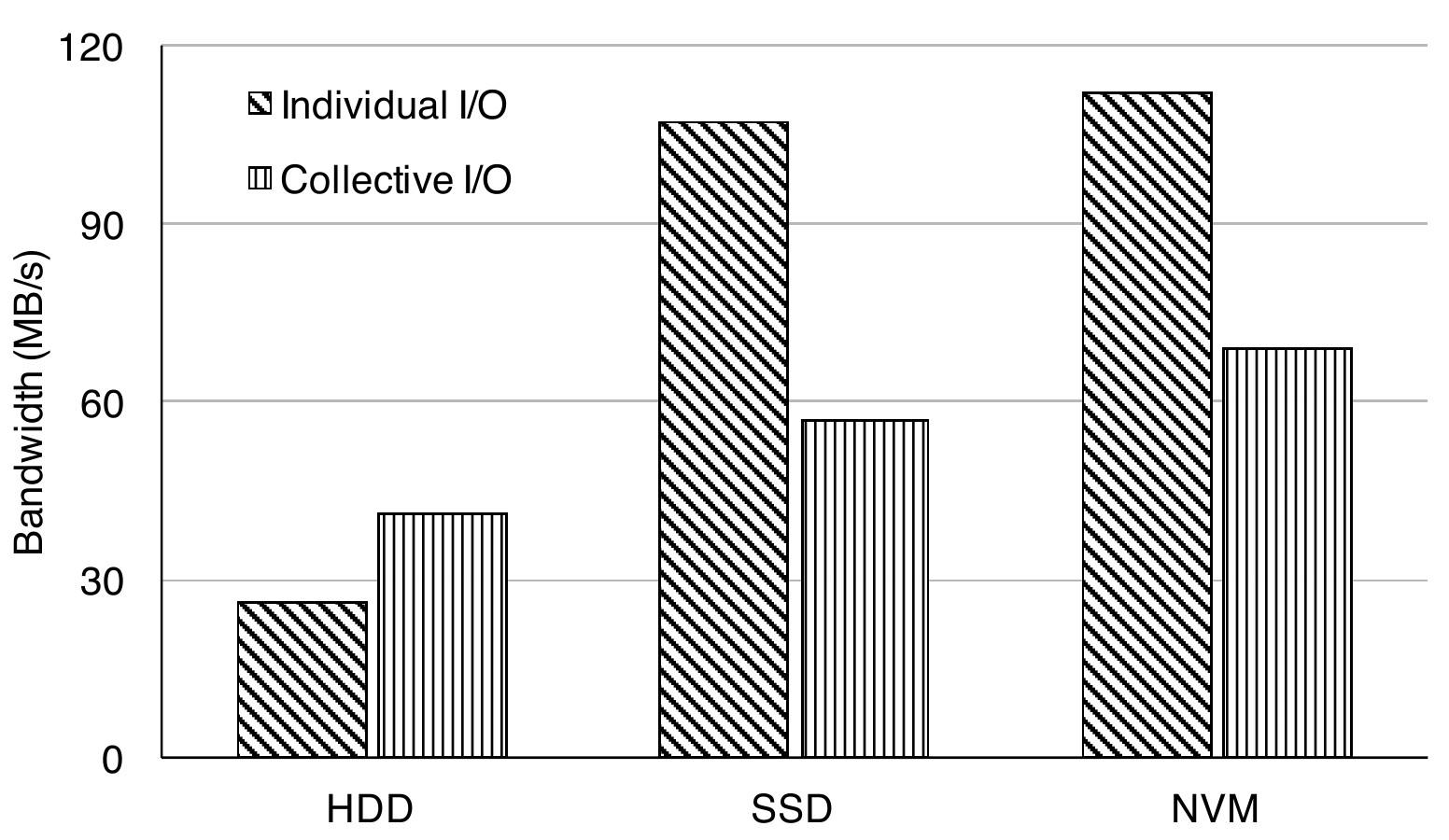}
\vspace{-5pt}
\caption{Comparing the performance of MPI collective I/O and MPI individual I/O (4 processes per node) with IOR.}
\vspace{-20pt}
\label{fig:IOR_IC16}
\end{figure}

We use the IOR benchmark and the same configuration (including workload size, block size, and data transfer size) as that for MPI individual I/O and POSIX I/O (Section~\ref{sec:posix_vs_individual}). 
%With our IOR configuration, the I/O operations are 
%(\textbf{TODO: describe how I/O operations happen on multiple processes. Are they interleaved?})
We use five nodes for the tests, four of which run IOR, and
the fifth node works as a remote storage node for parallel I/O operations.
For MPI collective I/O, we use one aggregator per node. 
Figures~\ref{fig:IOR_IC4} and~\ref{fig:IOR_IC16} show results for the case of 1 process per node (4 processes in total) and 4 processes per node (16 processes in total), respectively.

\begin{comment}
\begin{figure}
\centering
\includegraphics[width=0.46\textwidth, height=0.28\textheight]{figures/IC64.pdf}
\caption{IOR Individual I/O and Collective I/O Performance Difference (64 Processors)}
\label{fig:IOR_IC64}
\end{figure}
\end{comment}

%As for HDD, collective I/O indeed has a significant better performance than individual I/O. Which means, collective I/O indeed takes effect in reducing network overhead and improve the performance overall. While, situation in SSD and PMBD is totally different. In any cases, collective I/O has a much worse performance than individual I/O.

%One explanation about this phenomenon is NVM speed is very fast so that the time for data shuffling have increasing overhead ratio than the actually I/O latency. Then finally lead to collective I/O has a worse performance in the whole.

%After all, using collective I/O in NVM has a over all worse performance than just simple individual I/O. But how exactly data shuffle infect the performance and how fast the device will has the verse vise performance situation. We will continue our future study in next section.

The figures reveal that MPI collective I/O brings little benefit in most of cases.
For HDD with intensive I/O operations (i.e., 4 processes per node), 
the collective I/O performs better. 
But, for SSD and NVM, MPI collective I/O always performs worse than MPI individual I/O.
%The figures reveal that HDD benefits from the optimization of MPI collective I/O which performs better than MPI individual I/O. On the contrary, SSD and NVM achieve better performance with individual I/O. %than with MPI collective I/O. 
%We have similar observation with 64 processes on 4 nodes, which is more I/O intensive.

With conventional HDD, MPI collective I/O demonstrates its performance benefits, even if there is data shuffling cost.
However, with the introduction of faster storage device (e.g., SSD and NVM), the I/O cost on the storage device is alleviated, and relatively, the data shuffling cost becomes more pronounced in the overall I/O cost.
The results suggest that using MPI individual I/O instead of collective I/O makes more sense for fast storage device due to its low overhead.

Furthermore, we notice that the performance difference between
MPI collective I/O and individual I/O becomes bigger in the case of 4 processes per node than in the case of 1 process per node.
Such larger performance difference is due to the higher data shuffling cost when dealing with a large number of concurrently running processes.

\textbf{Conclusions.}
MPI I/O used to assume slow and pattern-sensitive HDDs as the secondary storage, which makes collective I/O a desirable optimization choice, disregarding the associated small overhead. As storage device performance improves to a point that the performance benefit cannot offset such overhead, MPI collective I/O becomes a detrimental ``optimization'', especially for NVM.
This urges us to revisit other existing mechanisms, besides MPI collective I/Os, that aim to optimize performance based on the ill assumption of slow storage devices.
With the emergence of NVM, the existing mechanisms may not be necessary and could be even harmful.
In this case, we demonstrate that MPI collective I/O is one of such mechanisms.

In Section~\ref{sec:perf_model}, we further study the performance of MPI collective I/O and investigate why it has worse performance. We also introduce a performance model that facilitates to make a decision on when to use MPI collective I/O.

\section{Detailed Performance Study for MPI Collective I/O}
\label{sec:perf_model}

%From the previous section, we found that collective I/O has certain overhead other than improvement in NVM. To understand how exactly it affects the performance, we continued our study about collective I/O by profiling out the different I/O phases and then we made a model to predict the performance variety.
MPI collective I/O is more than just I/O operations. It includes
a set of communication between participating MPI processes before or after
I/O operations. We conduct a detailed analysis on the performance of MPI collective I/O. %in this section. 

\subsection{Workflow of MPI Collective I/O}
%In collective I/O, each I/O iteration do the following works: As for read, aggregator first does IO phase in which data pieces will be read from storage and distribute into computing nodes. In data shuffle phase, aggregator will computes what exactly piece of data will be distributed and what process will take it. As for write, IO phase and shuffle phase will be versed from read, means data will be computed and merged in shuffle phase and then write back into storage node in IO phase. Listing~\ref{code:collective_read} and Listing~\ref{code:collective_write} shows the logic of main loop about collective read.
MPI collective I/O performs differently for read and write I/O operations. For read operations, the aggregator processes fetch data from the remote storage node and then redistribute the data among other MPI processes. For write operations, the aggregator processes collect data from other MPI processes and then write the data to the storage node. 
As discussed in Section~\ref{sec:bg}, the whole dataset is partitioned into many data blocks, and the aggregators scatter/gather data among MPI processes iteratively.

%Listings~\ref{code:collective_read} and ~\ref{code:collective_write} show the workflow for read and write operations in MPI collective I/O, based on the implementation of MPI collective I/O in ROMIO~\cite{romio99}.

Listing~\ref{code:collective_write} shows the workflow for write operations in MPI collective I/O, based on the implementation of MPI collective I/O in MPICH (in particular, ROMIO~\cite{romio99}).
%but the workflow for write operations is similar. 
%Base on the implementation of MPI collective I/O in ROMIO~\cite{romio99}, 
In each iteration of MPI collective I/O ($ntimes$ iterations in total), before each collective data write (Line 10), %and after each collective data write, 
data shuffling is called to gather data from MPI processes (Line 7).
%manipulate the possible interleaves between data pieces. 
%For either read or write operations, data shuffling and I/O operations are interleaved. 

Listing~\ref{code:shuffle} shows the logic of data shuffling in each iteration of MPI collective I/O.  Data shuffling is implemented based on 
%an MPI collective communication (particularly MPI\_Alltoall) and 
MPI asynchronized point-to-point communication (MPI\_Irecv/MPI\_Isend and MPI\_Waitall).

Based on the above discussion, we conclude that MPI collective I/O alternates between data shuffling and I/O operation.
%we conclude that data shuffling in MPI collective I/O is interleaved with IO operation across iterations of MPI collective I/O. 
In each iteration, data shuffling must be finished before the aggregator starts to write (or read) data. 
%Assuming the performance is dominated by the slowest aggregator, 
From the view of an individual aggregator,
data shuffling and I/O can be treated as blocking operations.
%, which simplifies our I/O modeling. 
%{ \textbf{Make sure you understand here, ask me, -Jialin}} [{\bf please make sure each code block showing completely, not broken into two columns --Feng}]

%In Listings~\ref{code:collective_read} and ~\ref{code:collective_write}, we have $ntimes$ of iterations and each iteration sends one data piece with the size of ``collective buffer''.

\begin{comment}
\begin{lstlisting}[language=c++,caption=Pseudocode for MPI collective read operations, label=code:collective_read, linewidth=9cm]
ADIOI_read_and_Exch(...)
{
  ...
  for (m=0; m<ntimes; m++) {
      ...
      // Contiguous read from storage
      ADIO_ReadContig(...);
      ...
      // Shuffling data between MPI processes
      ADIOI_R_Exchange_data(...);
      ...
  }
  for (m=ntimes; m<max_ntimes; m++) {
      // Nothing to send, but check for write.
      ADIOI_R_Exchange_data(...); basicstyle=\small
  }
  ...
}
\end{lstlisting}
\end{comment}

\begin{lstlisting}[language=c++,caption=Pseudocode for MPI collective I/O write operations, label=code:collective_write, linewidth=9cm, basicstyle=\footnotesize\ttfamily]
ADIOI_Exch_and_write(...)
{
  ...
  for (m=0; m<ntimes; m++) {
  	...
    // Shuffling data between MPI processes
	ADIOI_R_Exchange_data(...); 
    ...
    // Contiguous write to storage
    ADIO_WriteContig(...);
    ...
  }
  ...
}
\end{lstlisting}

%Between each collective I/O iteration, there’s no overlap between because each iteration has a barrier in the end. Like Table II shows, every iteration will be blocked until last iteration has been completed.

\begin{lstlisting}[language=c++,caption=Pseudocode for data shuffling in MPI collective I/O, label=code:shuffle, linewidth=9cm, basicstyle=\footnotesize\ttfamily]
ADIOI_R_Exchange_data(...)
{
  MPI_Alltoall(...);
  
  for (i=0; i < nprocs; i++) {
      MPI_Irecv(...)
  }
  
  for (i=0; i < nprocs; i++) {
      MPI_Isend(...)
  }
  
  MPI_Waitall(...)
  ADIOI_Fill_user_buffer(...)
  MPI_Waitall(...)
}
\end{lstlisting}
\vspace{-10pt}

\subsection{Profiling MPI Collective I/O}
\label{sec:profile}
Based on the above analysis on the implementation of MPI collective I/O, we add timers to measure the performance of data shuffling ($T_s$) and read/write ($T_{IO}$) operations in each iteration of MPI collective I/O. 
%Using the timers, we record performance of data shuffling in each iteration of MPI collective I/O. 
%we record the performance for each data piece (i.e., data exchanged during one iteration of the data shuffling loop).
%in Listings~\ref{code:collective_read} and~\ref{code:collective_write}). 
%Take this attribute, we used a timer to record each phase in every iteration. Average I/O time Tio will indicate device I/O speed, while average shuffle time Ts will indicate time spent in data shuffle. Tio will varies according to device speed, ratio of Ts/Tio will indicate how shuffle time impact in overall collective I/O.

%Let's firstly take a look at how much difference between I/O time and shuffle time in a big picture. We recording the I/O time and shuffle time by putting timers before and after I/O date exchange call and data shuffle call in MPICH source code. Time are stored in an array, average I/O time, average shuffle time, I/O time and shuffle ratio could be computed out in this case.

During profiling, we use the same five nodes as Section~\ref{sec:coll_io_perf_study}. Among the five nodes, four of them run IOR and one works as a storage node. For IOR, we use 16 processes (4 processes per node), and set ``segment count'', ``block size'', and ``transfer size'' 
as 2, 512MB, and 16MB respectively.
Total workload size for the four nodes is 16 GB.
%2 data segment counts, 512 MB block size. Total workload size sum up is 16GB. Transfer size for each collective iteration is 16MB. 
We use one aggregator per node.
Table~\ref{tab:IOR_profile} shows our profiling results. 
%Configuration for hardware and benchmark remains the same as overall performance evaluation in previous section. We used 1 main storage node connected with 4 remote computing nodes. Memory size is 12GB, workload size is 16GB, block size is 256MB, transfer size is 16MB. 16 processors are used here, so each computing node has 4 process. One aggregator for every computing node responses for data shuffle. 

\begin{comment}
\begin{table*}[!tpbh]
\footnotesize
     \begin{center}
     \caption{Profiling results for MPI collective I/O with IOR}
     \vspace{-5pt}
     \label{tab:IOR_profile}
     \begin{tabular}{|p{6cm}|p{2.2cm}|p{2.2cm}|p{2.2cm}|}
     \hline
     \textbf{Item}  & \textbf{HDD} &  \textbf{SSD} & \textbf{NVM}   \\ \hline
     I/O time (s)	& 5938.91	& 1002.93	& 986.15	\\\hline
     Shuffle time (s)	& 466.21	& 499.30	& 494.61	\\ \hline
     Ratio (shuffle time to collective I/O time) & 7.85\% & 49.93\%  & 50.16\%  \\ \hline
     Average IO time per iteration (ms)	& 170.38	& 28.77	& 28.29	\\ \hline
     Average shuffle time per iteration	(ms) & 13.77	& 14.32	& 14.19	\\  \hline
     %IO time variance	& 0.15	& 0.05	& 0.03	\\ \hline
     %Shuffle time variance & 0.03	& 0.02	& 0.02	\\ \hline
     %Ratio (shuffle time to IO time) & 49.13\%	& 89.74\%	& 93.48\%	\\ \hline
     \end{tabular}
     \vspace{-20pt}
     \end{center}
\end{table*}
\end{comment}

\begin{table}[!tpbh]
\scriptsize
     \begin{center}
     \caption{Profiling results for MPI collective I/O with IOR}
     \vspace{-5pt}
     \label{tab:IOR_profile}
     \begin{tabular}{|p{3.8cm}|p{1cm}|p{1cm}|p{1cm}|}
     \hline
     \textbf{Item}  & \textbf{HDD} &  \textbf{SSD} & \textbf{NVM}   \\ \hline
     I/O time (s)	& 5938.91	& 1002.93	& 986.15	\\\hline
     Shuffle time (s)	& 466.21	& 499.30	& 494.61	\\ \hline
     Ratio (shuffle time to collective I/O time) & 7.85\% & 49.93\%  & 50.16\%  \\ \hline
     Average IO time per iteration (ms)	& 170.38	& 28.77	& 28.29	\\ \hline
     Average shuffle time per iteration	(ms) & 13.77	& 14.32	& 14.19	\\  \hline
     %IO time variance	& 0.15	& 0.05	& 0.03	\\ \hline
     %Shuffle time variance & 0.03	& 0.02	& 0.02	\\ \hline
     %Ratio (shuffle time to IO time) & 49.13\%	& 89.74\%	& 93.48\%	\\ \hline
     \end{tabular}
     \vspace{-20pt}
     \end{center}
\end{table}

The table reveals that from HDD, SSD, to NVM, the ratio of shuffle time to total collective I/O time increases from 7.85\% to 50.16\%.
The shuffle time accounts for a larger percentage of performance loss,
when we use NVM. %as the storage device. 
Note that the shuffle time remains stable across the cases of HDD, SSD, and NVM,
even through the ratio is different in the three cases. 
Because we use the same MPI implementation and the same I/O workload  for the three cases, the communication pattern 
should be identical for the three cases and the shuffle time should be stable across the three cases.
%which results in identical communication pattern for the three test cases. 
%From Table~\ref{tab:IOR_profile} we could have an overall view about what the consistent of collective I/O. Firstly, I/O times for three devices have certain difference. NVM quicker than HDD and SSD, so I/O time is less. We have to point out, the difference has been flatten by network overhead. Then, the shuffle time among any devices are the same. The reason is easy to understand, that because all data shuffle work is done in memory, nothing related to storage devices. In this case, from HDD to SSD and NVM, I/O time is decreasing while shuffle time remain consistent, shuffle time to dominate the overhead in collective I/O.

%An additional find out is NVM I/O time variance is 80 percentage less than HDD, 40 percentage less than SSD, means NVM is more consistence in I/O performance. 

%To continue our study, we want to find out in what circumstance the collective I/O will have an overhead, other than improvement. We set an guess linear regression model according the collective I/O mechanism.

\subsection{Performance Modeling for MPI Collective I/O}

We model MPI collective I/O performance based on the above discussion.
The notation for our models is summarized in Table~\ref{tab:model}.
%Let's firstly look at collective I/O. Because collective read and write are pretty much the same, (the only difference is the order of data exchange and data shuffle), we take collective read as example now. Please refer to Table~\ref{tab:args} for arguments explanation.

\begin{table}
\footnotesize
     \begin{center}
     \caption{Notation of our performance modeling for MPI collective I/O}
     \vspace{-5pt}
     \label{tab:model}
     \begin{tabular}{|p{1.6cm}|p{5.4cm}|}
     \hline
     $T_{collective}$	& The collective IO time. 	\\\hline
     $T_{individual}$	& The individual I/O time. 	\\\hline
     $T_{comm}$	& Data shuffling time. 	\\\hline
     $T_{IO}$	& IO operation time. 	\\\hline
     $T_{other}$	& Other performance cost besides data shuffling. 	\\\hline
     %$tran\_size$	& Transfer size for each iteration. 	\\\hline
     $msg\_size_i$	& The size of data that are communicated between the slowest aggregator and each MPI process for data shuffling in an iteration $i$. \\\hline
%The collective buffer size in the collective I/O implementation. 	
     $\tau$	& The ratio of data participated in data shuffling to total data. 	\\\hline
     $iter$	& The number of iterations within the iterative collective I/O. 	\\\hline
     $T_{w}$	& Communication time independent of the message size.   	\\\hline
     $T_{s}$	& Communication time in proportion to the message size 	\\\hline
     %$BLK$	& Block size. 	\\\hline
     %$a$	& Collective aggregator number.  	\\\hline
     $bdw_{seq}$	& Sequential end-to-end I/O bandwidth.  	\\\hline
     $bdw_{ran}$	& Random end-to-end I/O bandwidth.  	\\\hline
     \end{tabular}
     \vspace{-15pt}
     \end{center}
\end{table}

%From Listing~\ref{code:collective_write}, the aggregator mainly just do two things: data shuffle and data exchange, we used $T_{comm}$ and $T_{io}$ to represent them. There are also other things aggregator has to do, for example function stack call, inter and intra process communication, we take this time as $T_{other}$. So we have equation~ref{eqal:coll} indicates the time for collective I/O: $T_{collective}$.

\textbf{MPI collective I/O} ($T_{collective}$) is generally modeled in Equation~\ref{eq:general}.
The equation includes the data shuffling time ($T_{comm}$),
I/O operation time ($T_{IO}$), and other performance cost ($T_{other}$) because
of the implementation of MPI collective I/O.
$T_{comm}$ and $T_{IO}$ depend on data size and data access patterns of MPI processes. We model them as follows.

\begin{equation}
\small
\label{eq:general}
T_{collective} = T_{comm} + T_{IO} + T_{other}
\end{equation}

%As for data shuffle time: $T_{comm}$. We set a linear regression to train it as formula below. Several arguments explain here: $iter$ is the iteration time in whole test. One test is divided by several processors, aggregators and buffer containers. So, the time for each iteration multiple the iteration times $iter$ equals to $T_{comm}$. MSG is the work load size. For this experiment, MSG is 16 GB. $\tau$ is the ratio of how much data pieces will do collective I/O computed by aggregator. This ratio is depends on the certain benchmark. For our IOR test, $\tau$ is 1.0 because we had set random read back flag and profile log indicates all iterations have data shuffle work. $T_{w}$ and $T_{s}$ is the slope and intercept of linear regression representatively. $T_{w}$ also represent the bandwidth of each processor and $T_{s}$ means all other works in an iteration.

Data shuffling time ($T_{comm}$) is modeled in Equation~\ref{eq:t_comm}. 
$T_{comm}$ is for one MPI aggregator (the slowest aggregator). There might be multiple aggregators involved in the collective I/O, but their data shuffling times are overlapped.
The data shuffling phase iteratively sends or receives data between the aggregator and other MPI processes. 
%\textbf{those MPI processes that are not an aggregator, but are in the same node as the aggregator}. 
%In each collective I/O iteration, how much data be communicated is defined by transfer size--``tran\_size''. 
%\textbf{TODO: verify if the above statement on $iter$ is correct.}

In Equation~\ref{eq:t_comm}, at a specific iteration $i$,  $msg\_size_i$ of data is communicated between the aggregator and each MPI process for data shuffling.
In total, $\sum_{i=1}^{iter}msg\_size_i$ of data, which is the total amount of data from one MPI process for doing I/O operation, is communicated. There might be multiple MPI processes concurrently communicating with the aggregator shown in Lines 6 and 10 of Listing 2,  but their communication times are overlapped. 
%Note that even though $msg\_size$ is the data the MPI process needs to be communicated for I/O operations. The real message size between a MPI process and the aggregator can be smaller than $msg\_size$, 
Note that %Equation~\ref{eq:t_comm} is only for modeling collective I/O, in which
it is possible that only a part of the total data is really communicated, while the other part of the data already reside in some aggregator and do not need to be transfered between the aggregators.
%in some I/O aggregator and do not need to be transfered between it and the slowest aggregator.
%because the aggregator can choose to gather and scatter a part of the data, if the other partial data is already contiguous and friendly for doing I/O operations individually. 
To capture the above fact, 
we introduce a parameter, $\tau$. So, $msg\_size_i \times \tau$ is the amount of data that is really
involved in the data shuffling between an MPI process and the slowest aggregator. %[\textbf{Corrected, Jialin}] 
Note that $\tau$ is application-dependent and related with the application's inherent I/O access pattern.
%\textbf{TODO: verify if the above statement on $\tau$ is correct.}

%Once the real message size for communication is determined, 
Based on the above discussion,
the communication time for an iteration $i$ is modeled by $T_s + T_w \times msg\_size_i \times \tau$, in which
$T_s$ represents the communication time unrelated with the message size, such as communication initialization time, and 
$T_w$ represents the communication time related with the message size (or more precisely speaking, in proportion to the message size).

\begin{comment}
Because data shuffle happens in parallel among computing nodes, the real shuffle time is sum up of all collective iteration shuffle time.
%and divided by node number ($m$).\textbf{why divided by m, -Jialin}
Equation~\ref{eq:t_comm} summarize the model to compute the shuffle time.
\end{comment}

\begin{equation}
\small
\label{eq:t_comm}
%T_{comm} = (T_s + T_w \times tran\_size \times \tau) \times iter
T_{comm} = \sum_{i=1}^{iter} (T_s + T_w \times msg\_size_i \times \tau) 
\end{equation}

%For workload size: $msg\_size$. It will divided by several computing node number $m$ and aggregators a. After all, we still need to multiple the number of aggregator a to get the total size for each iteration. So, $msg\_size$ is actually a formula related to workload size and computing node number $m$. 

%\begin{equation}
%\label{eq:mg_size}
%msg\_size = \frac{\sum_{i=1}^{n}iBLKi}{a \times m} \times a = \frac{\sum_{i=1}^{n}iBLKi}{m}
%\end{equation}

I/O operation time ($T_{IO}$) is modeled in Equation~\ref{eq:t_io_coll}.
The numerator of the equation is the data  %has been shuffled and 
ready for I/O operation.
%as discussed above in $T_{comm}$. %In this equation, 
$bdw_{seq}$ in the denominator is the end-to-end bandwidth (between the end of a compute node and the end of a storage node), and $bdw_{seq}$ is the bandwidth for doing sequential I/O, because after data shuffling, there is supposed to be sequential data accesses between the aggregator and storage node. 
%differs among different storage devices. As all basic I/O manipulation, exchange time is determined by data size and the bandwidth ($bdw$) of the device working on~\ref{eq:t_io_coll}. Total workload size times collective I/O ratio ($\tau$) is the total I/O size for this case.

\begin{equation}
\label{eq:t_io_coll}
\small
T_{IO} = \frac{\sum_{i=1}^{iter}msg\_size_i}{bdw_{seq}}
%T_{IO} = \frac{\sum_{i=1}^{iter}(msg\_size_i \times \tau)}{bdw_{seq}}
\end{equation}
%(\textbf{removed $\tau$, Jialin})

$T_{other}$ in Equation~\ref{eq:general} is the other performance cost besides data shuffling, including memory mapping, variable initialization, system logs, and data checking for data alignment.
%(See Lines 13-15 in Listing~\ref{code:collective_read} and Listing~\ref{code:collective_write}).
%Other time ($T_{other}$) is the other performance cost, including memory mapping, variable initialization, system logs, and data checking for data alignment (See Lines 13-15 in Listing~\ref{code:collective_read} and Listing~\ref{code:collective_write}).
%\textbf{TODO: explain what is bdw\_1 and how to get bdw\_1.}
%\textbf{TODO: finish the above statement}.
%We assume $T_{other}$ is constant, and independent of the data size for I/O operations. We can easily measure $T_{other}$ with a micro-benchmark doing I/O operation with a tiny data size (e.g., 100 byte).
%\textbf{TODO: how $T_{other}$ can be measured is questionable. Change the above statement.}
%\textbf{LIU: we don't have experiment to measure $T_{other}$ right now.}

\textbf{MPI individual I/O.} 
To make a comparison between MPI collective I/O and individual I/O, we also model the performance of individual I/O, shown in Equation~\ref{eq:t_io_indv_total}.
$T_{individual}$ is much simpler than the collective I/O, because it does not have data shuffling, and I/O operations ($T_{IO}$) from each MPI process happen independently. 
To calculate $T_{IO}$, we use the end-to-end bandwidth for random data access ($bdw_{ran}$), shown in Equation~\ref{eq:t_io_indv}. 
This is based on an assumption that data accesses from MPI processes are random without coordination as the collective I/O, but whether this assumption is true depends on the data access pattern of the application.
%($T_{individual}$) is much simpler than collective I/O. Because it simply does the data exchange, and  $T_{comm}$ in collective I/O formula~\ref{eq:t_comm} need to be crossed out. Besides, $T_{comm}$ is the workload size divided by device bandwidth, because all iterations will do data exchange.
%\textbf{TODO: explain what is bdw\_2 and how to get bdw\_2.}

\begin{equation}
\small
\label{eq:t_io_indv_total}
T_{individual} = T_{IO} + T_{other}
\end{equation}

\begin{equation}
\small
\label{eq:t_io_indv}
T_{IO} = \frac{\sum_{i=1}^{iter}msg\_size_i}{bdw_{ran}}
\end{equation}

\textbf{Model usage.}
To use the model, we need to know a set of parameters, including application-independent ones and application-dependent ones.
The application-independent parameters include $T_s$, $T_w$, $bdw_{seq}$, $bdw_{ran}$, and $T_{other}$, which are measured only once on any platform.
The application-dependent parameters include $msg\_size$, $\tau$, and number of iterations $iter$.

$T_s$ and $T_w$ are measured by running an MPI-based micro-benchmark doing ping-pong communication between compute node and storage node with different message sizes.
We measure the communication time for each message size and use a linear regression to get $T_s$ and $T_w$. In our test environment, $T_s = 5.39e-3$ (s) and $T_w = 3.35e-2$ (s/MB).

The parameters, $bdw_{seq}$ and $bdw_{ran}$, can be measured by using IOR.
In particular, we deploy IOR on our test environment with four compute nodes
and one storage node. Using IOR, we perform read or write I/O operations
for 2GB data. We set ``reorder tasks to random'' to enable either random
or sequential I/O accesses with 16 MPI processes (4 processes per node), and then calculate $bdw_{seq}$ and $bdw_{ran}$.
Table~\ref{tab:meas_bdw} summarizes the results in our test platform.
One interesting observation is that between SSD and NVM, there is no big difference 
in terms of $bdw_{seq}$ and $bdw_{ran}$, shown in the table.
This is because of the fact that SSD and NVM have a larger device bandwidth than HDD, such that the end-to-end bandwidth is limited by networking. 

\begin{comment}
IOR write and read for 2GB data using the same multiple nodes hierarchy we used for our profiling benchmark (4 remote computing nodes connected with one storage node). Set the read back order as ``sequential'' or ``random'' to control the I/O pattern as needed. Table~\ref{tab:meas_bdw} shows the results. Note that due to the network overhead, SSD and NVM do not show substantial performance difference through multiple nodes.
\end{comment}

\begin{table}[h]
\footnotesize
     \begin{center}
     \caption{$bdw_{seq}$ and $bdw_{ran}$ in our test platform.}
     \vspace{-5pt}
     \label{tab:meas_bdw}
     \begin{tabular}{|p{2.8cm}|p{1.2cm}|p{1.2cm}|p{1.2cm}|}
     \hline
     	& \textbf{HDD} & \textbf{SSD} & \textbf{NVM}	\\\hline
     $bdw_{seq}$ (MB/s)	& 58.11 & 110.98 & 112.31	\\\hline
     $bdw_{ran}$ (MB/s)	& 26.72 & 101.86 & 110.51	\\\hline
     \end{tabular}
     \vspace{-15pt}
     \end{center}
\end{table}

$T_{other}$ is assumed to be constant in our model, and can be measured through IOR as well. In particular, we deploy the same tests as the ones for measuring $bdw_{seq}$ and $bdw_{ran}$, and measure $T_{individual}$, $T_{collective}$, 
I/O operation time and shuffling time. Then, we calculate $T_{other}$
based on Equations~\ref{eq:general} and~\ref{eq:t_io_indv_total}
for collective I/O and individual I/O, respectively.
In our tests, we find that $T_{other}$ is much smaller than I/O operation time
and data shuffling time. Hence we set $T_{other}$  as zero during model validation (Section~\ref{sec:model_verif}).

The total data size (see the discussion on $\sum_{i=1}^{iter}msg\_size_i$ in the MPI collective I/O modeling) can be obtained by examining the application, particularly MPI I/O calls (e.g., MPI\_File\_write\_all() and MPI\_File\_read\_all()).
In each iteration, $msg\_size$ is constant in our model, which is equal
to the collective buffer size (16MB in our tests) in ROMIO. The number of iterations ($iter$) is equal to the total data size divided by the constant collective buffer size. %(in practice, it is the minimum of $msg\_size$ and buffer size).
%(\textbf{Corrected, Jialin})

The parameter $\tau$ depends on the application I/O access pattern and MPI implementation. It is challenging to predict or choose a universally appropriate value for all possible cases.
Also, it is challenging to ask the user to quantify their workload characteristics
and choose $\tau$. For simplicity,
assuming that each MPI process needs to do the same size of IO, and during the two phases (data shuffling and I/O phases) of collective IO, the data sent from all non-aggregator MPI processes are evenly handled by the aggregators, then we roughly estimate $\tau$ as the number of non-aggregator MPI processes divided by total number of MPI processes. 

For example, suppose that we have 8 processes in total, 2 of them are I/O aggregators, and each process needs to write 1MB data (i.e., 8MB for 8 processes). Then during the two phases of collective IO, the 6 non-aggregators need to send totally 6MB data to the 2 aggregators. 
Based on the definition of $\tau$, $\tau$ = 6MB/8MB = $75\%$.
Based on our estimation of $\tau$, $\tau =$ (\#non-aggregators / total number of MPI processes) = $6/8 = 75\%$.
Our estimation of $\tau$ has a great match to the real value of $\tau$.
%In this example, $\tau=6/8=75\%$, which is the actual communication ratio. 

Note that when estimating $\tau$, we assume that the data sent from all non-aggregator MPI processes are evenly handled by the aggregators.
In practice, the data can be unevenly handled by the aggregators.
It is even possible that some aggregator does not need to do any data shuffling.
However, our model is for the slowest aggregator that has the longest
data shuffling time and dominates data shuffling time of all aggregators.
$\tau$ for the slowest aggregator can be estimated well by our method in most cases.
%the communication load on each aggregator can be different. It is possible that some aggregator does not need to do any communication.
 
%Note that our model is for one aggregator, which is assumed to be the slowest dominant aggregator, the actual communicated data size on this aggregator is hard to predict. We use the estimated average communication ratio for $\tau$ is based on the assumption that the workload is evenly handled by the selected aggregators in current ROMIO implementation. In practice, the communication load on each aggregator can be different. (It is possible that some aggregator does not need to do any communication)
%(\textbf{Jialin, Rewritten}) 

\textbf{Discussion.}
Our model has two limitations. First, we do not distinguish 
intra- and inter-node communication in Equation~\ref{eq:t_comm} when modeling data shuffling time.
In particular, we measure $T_s$ and $T_w$ based on inter-node communication and use them in Equation~\ref{eq:t_comm}, no matter whether data shuffling happens within a node or between nodes. 
Second, we assume that 
%all aggregators have roughly the same data shuffling time, such that 
the data shuffling times of all aggregators can be greatly overlapped. 
However, depending on data access patterns of MPI processes, 
different aggregators working with different MPI processes can have 
different, non-overlapped data shuffling time.
%No communication and I/O overlap;

To fix the above model limitation, we must have good knowledge on the execution environment, such that we know how MPI processes are mapped to nodes in order to determine intra- and inter-node communication; we must also have deep knowledge on data access patterns of each MPI process. 
However, having the above knowledge greatly limits the model usability
and generality, while providing limited help for modeling accuracy.
Hence, we do not assume such knowledge is available in our model. Our results show that the current model works reasonably well. 
%[{\bf Added one sentence here. -Feng}]

\subsection{Model Validation}
\label{sec:model_verif}
We verify our model accuracy with IOR.
\begin{comment}
%on five nodes, four of which are compute nodes and one of which is a storage node, the same as our previous multi-node evaluation. We use four MPI processes per node (16 MPI processes in total and one aggregator per node). With IOR, 
We evaluate two types of workloads. One type of workload has random but overlapped data accesses from MPI processes. This is achieved by enabling ``reorder tasks to random''.
%For this type of workload, $\tau$ is set as 1 to predict collective I/O performance. 
The other type of workload has sequential, non-overlapped data access pattern for each MPI process.
For the first type of workload, we use five nodes, four of which are compute nodes (4 MPI processes per node) and one of which is a storage node, the same as our previous multi-node evaluation. 
For the second type of workload, we use three nodes, two of which are compute nodes (8 MPI processes per node) and one of which is a storage node. 
%We use four MPI processes per node (16 MPI processes in total and one aggregator per node)
%For this type of workload, $\tau$ is set as 0.2 to predict collective I/O performance. This also indicates that 20\% of data for I/O operations are based on the model for collective I/O, while 80\% of the data for I/O operations are based on the model for individual I/O. 
\end{comment}
We test two cases, one with 4 compute nodes (4 processes per node) and the other with 2 compute nodes (8 processes per node). Both cases have one storage node. 
For IOR, ``segment count'', ``block size'', and the collective buffer size are set as 2, 64MB, and 16MB respectively. We use one aggregator per node in validation tests.
%block size, 4 processors. We set collective buffer size ($msg\_size$) to 4MB, 8MB, 16MB, 64MB

Tables~\ref{tab:cpr_4_nodes} and~\ref{tab:cpr_4_nodes} show the validation results. In general, our model achieves high accuracy
in 12 validation tests (average error 4.93\% and at most 14.4\%). More importantly, our model correctly captures performance trend across the three devices in different cases.

\begin{table}
     \begin{center}
     \caption{Comparison of estimated and measured I/O times with 4 compute nodes (4 processes per node). The percentage numbers in brackets are prediction errors.}
     \vspace{-5pt}
     \label{tab:cpr_4_nodes}
     \scriptsize
     \begin{tabular}{|p{2cm}|p{1.4cm}|p{1.4cm}|p{1.4cm}|}
     \hline
     \textbf{Device}	& \textbf{HDD} & \textbf{SSD} & \textbf{NVM}	\\\hline
     \textbf{Collective I/O estimated time (s)}	& 411.78(6.7\%) & 286.21(3.2\%) & 284.46(14.4\%) 	\\\hline
     \textbf{Collective I/O Measured time (s)}	& 385.86 & 277.46 & 242.54	\\\hline \hline
          \textbf{Individual I/O estimated time (s)}	& 613.17(3.4\%) & 160.84(9.8\%) & 145.88(3.2\%) 	\\\hline
     \textbf{Individual I/O Measured time (s)}	& 593.04 & 146.50 & 146.35	\\\hline 
     \end{tabular}
     \vspace{-10pt}
     \end{center}
\end{table}

\begin{table}
     \begin{center}
     \caption{Comparison of estimated and measured I/O times with 2 compute nodes (8 processes per node). The percentage numbers in brackets are prediction errors.}
     \vspace{-5pt}
     \label{tab:cpr_2_nodes}
     \scriptsize
     \begin{tabular}{|p{2cm}|p{1.4cm}|p{1.4cm}|p{1.4cm}|}
     \hline
     \textbf{Device}	& \textbf{HDD} & \textbf{SSD} & \textbf{NVM}	\\\hline
     \textbf{Collective I/O estimated time (s)}	& 350.90(1.04\%) & 216.58(0.53\%) & 214.83(0.85\%) 	\\\hline
     \textbf{Collective I/O Measured time (s)}	& 354.59 & 217.74 & 213.01	\\\hline \hline
          \textbf{Individual I/O estimated time (s)}	& 613.17(5.66\%) & 160.84(9.86\%) & 145.88(0.46\%) 	\\\hline
     \textbf{Individual I/O Measured time (s)}	& 580.32 & 146.40 & 146.55	\\\hline 
     \end{tabular}
     \vspace{-15pt}
     \end{center}
\end{table}

\subsection{Model Implication}
\label{sec:model_implication}
%Now we have made a model and verified its accuracy. This collective I/O could give us a hint about when and how we use collective I/O in the future works. Comparing the two models about collective I/O and individual I/O, the factor to determine whether collective I/O performance better or individual I/O is that whether the benefit of doing the continuous read and write from aggregators is overcome the overhead of doing data shuffle. We could simply calculate out both the benefit from individual I/O to collective I/O of one data piece and the cost of shuffle that. Figure~\ref{fig:C_all} shows the comparison of the ``benefit'' and the ``cost'' by using different size data pieces.
Our model enables us to explore the tradeoff between data shuffling cost and collective I/O benefit in a variety of environments with different storage devices. Hence it can be used to enable adaptive performance optimization and improve I/O performance for the future HPC using NVM-based storage.

\begin{figure}
\centering
\includegraphics[width=0.48\textwidth, height=0.18\textheight]{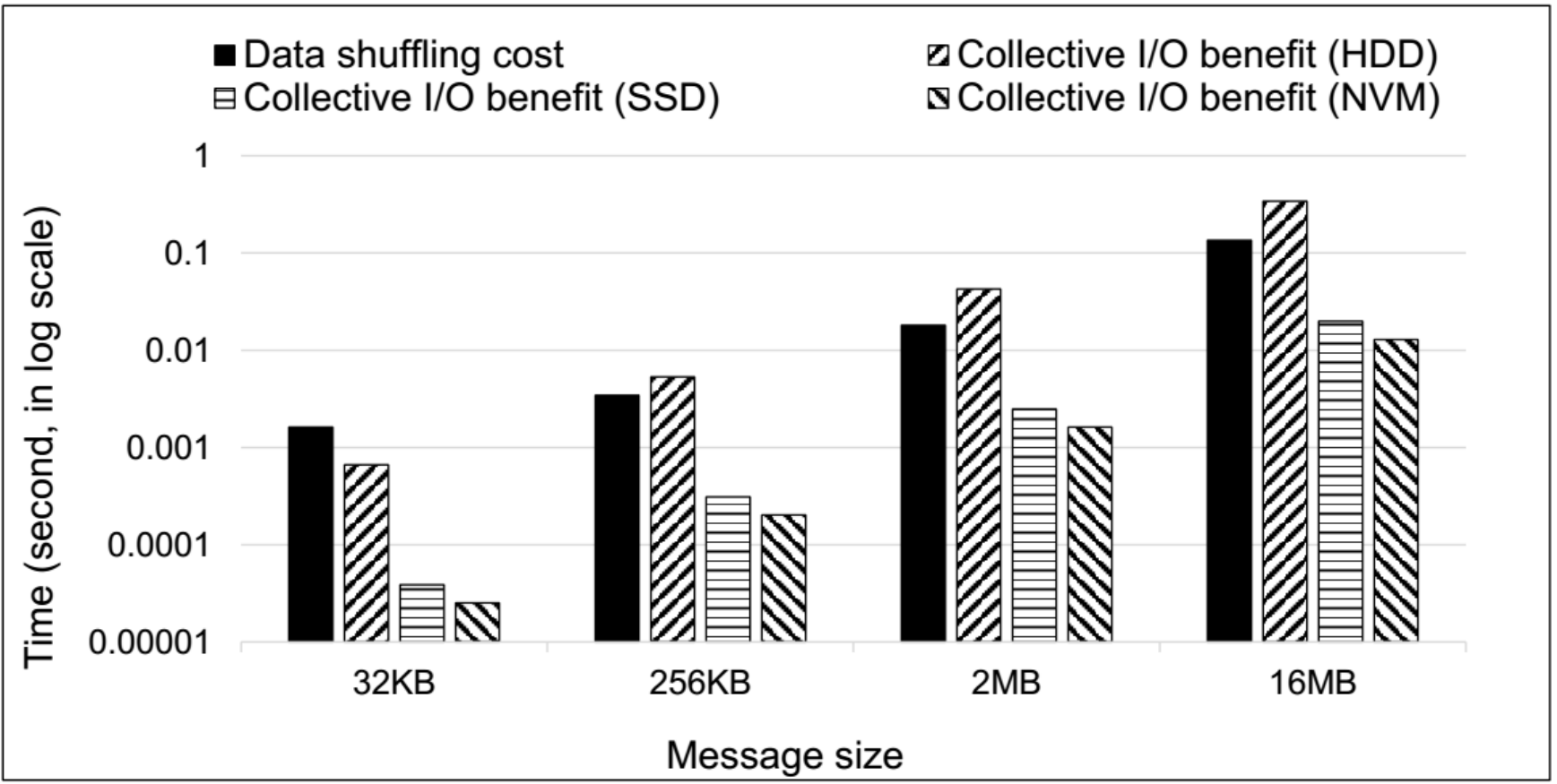}
\caption{Explore the performance tradeoff between data shuffling cost and collective I/O benefit.}
\vspace{-20pt}
\label{fig:C_all}
\end{figure}

As a case study, we use our model to study the tradeoff between data shuffling cost (Equation~\ref{eq:t_comm}) and collective I/O benefit. The collective I/O benefit is quantified by ($T_{individual} - T_{collective}$). 
We focus on one iteration (i.e., $iter=1$) and change the message size. We use bandwidth and communication parameters (i.e., $T_w$ and $T_s$) measured in our platform for our study. Figure~\ref{fig:C_all} shows the result, assuming that there are 4 compute nodes, 1 storage node, and 4 MPI processes per node.

The figure reveals that both data shuffling cost and collective I/O benefit increase as the message size increases, but at different rates. 
%The data shuffling cost increases much faster than the benefit.
For HDD, although the data shuffling cost is larger than the benefit when the message size is small (32KB), the data shuffling cost is smaller than the benefit when the message size is large (2MB and 16MB). However, for SSD and NVM, the data shuffling cost is always larger than the benefit, which explains why collective I/O performs consistently worse than individual I/O in Tables~\ref{tab:cpr_4_nodes}. %and~\ref{tab:cpr_2_nodes}. 
%\textbf{TODO: fix the above Table numbers.}

%As for HDD, when $msg\_size$ is greater than a certain value (e.g. around 100KB in our environment), doing the collective does have a better benefit than the cost of data shuffle; while when $msg\_size$ is less than this point, the cost of data shuffle will overcome the benefit of doing collective I/O. That's why collective I/O has a worse performance in small data pieces communication~\cite{reference:topo}. However, as for SSD and NVM, in any situation the cost of data shuffle always has a huge overhead than the benefit. Which means, NVM is not suit for collective I/O in any predictable circumstance. Avoiding using this mechanism will both boost performance as well as simplify our programming model in this device.

\section{Related Work}
\label{sec:related_work}

\textbf{Non-volatile memory.}
%Non-volatile memory (NVM) technology is under quick development and has attracted a large body of research. A comprehensive survey about NVM can be found in a prior study~\cite{NVMDB}. Here we summarize the most related prior work on NVM.
Prior NVM studies can be roughly classified into several categories. Some earlier studies focus on the architecture-level design issues of NVM~\cite{Lee09, Qureshi09MICRO, Qureshi09ISCA, Zhou09}, such as wear-leveling, read-write disparity issues, etc. Most of these studies consider NVM as a displacement of DRAM at the architecture level. Another alternative is to consider NVM as a storage device, such as Onyx~\cite{Akel11}, Moneta~\cite{Caulfield10}, and PMBD~\cite{MSST14FC}. The recently announced Intel Optane product~\cite{Optane} also falls into this category. Researchers have also studied on the system and application level support for NVM. Some prior studies have explored file systems for NVM. For example, BPFS~\cite{Condit09} uses shadow paging techniques for fast and reliable updates to critical file system metadata structures. SCMFS~\cite{Wu11} adopts a scheme similar to page table in memory management for file management in NVM. PMFS~\cite{Dulloor14} allows to use memory mapping (mmap) for directly accessing NVM space and avoids redundant data copies. In order to take advantage of byte-addressability and persistency of NVM, a large body of research on NVM is on developing new programming models for NVM. For example, Mnemosyne~\cite{Volos11} gives a simple programming interface for NVM, such as declaring non-volatile data objects. CDDS~\cite{Venkataraman11} attempts to provide consistent and durable data structures. NV-Heaps~\cite{Coburn11} gives a simple model with support of transactional semantics. SoftPM~\cite{Guerra12} offers a memory abstraction similar to~\verb+malloc+ for allocating objects in NVM. In this study we treat NVM as a storage device and deploy conventional file systems atop for HPC applications. Our observations have confirmed that the high-speed NVM could significantly improve HPC application performance, however, the end-to-end effect is  workload dependent and related to a variety of factors in the  I/O stack. %from application, MPI library, OS page cache, file system to NVM device. 

\textbf{MPI I/O.}
%Aside for hardware, there are also tons of studies related to MPI talks about MPI-IO, collective I/O and software pattern with a focus on improving computing performance, guarantee data safety, and simplify coding strategy.
ROMIO~\cite{reference:usingMPI} is a widely used implementation of MPI-IO, which is included in the MPICH library. ROMIO uses two-phase I/O strategy~\cite{reference:topo} for collective I/O. 
%This technology could reduce latency of access and improve its scalability. 
Some prior work explores the determination of 
%It is still an open topic about determining an 
optimal number of aggregators for MPI I/O~\cite{reference:automatselect}. 
%In this paper, we mainly used one aggregator and used multiple aggregators as comparison. Because no clear difference is shown in experiments about aggregator number, we leave further experimental studies on this issue as our future work.
%Several studies have evaluated application efficiency on MPI and tried to improve it. 
Some prior work 
takes into account the network topology for deciding aggregators and introduces an optimized buffering system to reduce the aggregation cost~\cite{reference:topo}.
%introduces an optimized buffering system to reduce the aggregation cost~\cite{reference:topo}.
%proposed an optimized buffering system in order to reduce the aggregation cost, %as so improving reading and writing data efficiency~\cite{reference:topo}. 
%Another study finds that the nature of collective I/O can have a negative impact on underlying caching algorithm, leading to unnecessary cache misses.
%Hence An evolution of access pattern proposed address this issue and results in a new collective I/O  aware cache management methodology~\cite{reference:accessPattern}. 
Another study performs collective I/O while retaining access
patterns of MPI processes before collective I/O to enable better cache management~\cite{reference:collectiveComputing}. Our work is different from the prior studies by considering the impact of NVM on MPI I/O.
%tried to further reduce the communication costs in collective I/O of multi-core cluster systems with non-exclusive scheduling by regulate the node sequence ~\cite{reference:reduceColl}.

\section{Conclusions}
We study the impact of upcoming NVM on HPC I/O.
Given distinct performance characteristics of NVM, the existing
I/O stack must be re-examined to optimize performance.
Through our comprehensive performance study and modeling, we reveal the diminishing benefits of page cache, ignorable overhead of MPI individual I/O, and inappropriate performance optimization of MPI collective I/O.
Our work lays some foundation for the deployment of NVM in the future HPC.

% has evaluated the HPC I/O performance based on different storage backend, i.e., HDD, SSD and PMBD. We set experiments by using scientific application benchmarks to test the impact of page cache, performance of POSIX and MPIIO library with NVM. Then we further profile the collective I/O in ROMIO, which is a major optimization in MPI and HPC I/O. Based on the experimental result, we have developed a model and verified its accuracy. This is a comprehensive study of NVM and its impact to HPC I/O software stack. In conclusion, we find that the I/O performance with NVM is not as sensitive to page cache as HDD and SSD are in many cases, so we could shortcut the memory consumption in page cache, improving cost efficiency. We have also confirmed that collective I/O benefit diminishes as the storage becomes faster. So a simplified I/O strategies should be designed and revisited in the era of exascale computing with non-volatile memory. 
%\textbf{Rephased, -Jialin}

\bibliographystyle{IEEEtran}
\bibliography{li,liuwei,feng}

\end{document}